\documentclass[useAMS,usenatbib,usegraphicx]{mn2e}  
\usepackage[usenames]{color}

\newcommand{\be}{\begin{equation}}
\newcommand{\ee}{\end{equation}}
\newcommand{\bea}{\begin{eqnarray}}
\newcommand{\eea}{\end{eqnarray}}

\title[WSLAPing A1689] 
{A Free-Form Lensing Grid Solution for A1689 with New Mutiple Images}  

\author[Diego et al.]  
  {Jose M. Diego$^1$\footnote{jdiego@ifca.unican.es}, 
   T. Broadhurst$^{2,3}$, 
   N. Benitez$^{4}$,
   K. Umetsu$^{5}$, 
   D. Coe$^{6}$, 
   I. Sendra$^{2}$,
  \newauthor
   M. Sereno$^{7}$, 
   L. Izzo$^8$, 
   G. Covone$^{9,10}$\\
$^{1}$IFCA, Instituto de F\'isica de Cantabria (UC-CSIC), Av. de Los Castros 
s/n, 39005 Santander, Spain\\
$^{2}$Fisika Teorikoa, Zientzia eta Teknologia Fakultatea, Euskal Herriko 
Unibertsitatea UPV/EHU\\ 
$^{3}$IKERBASQUE, Basque Foundation for Science, Alameda Urquijo, 36-5 48008 
Bilbao Spain\\
$^{4}$Instituto de Astrof\'isica de Andaluc\'ia (CSIC), Apdo. 3044, 
      18008 Granada, Spain\\
$^{5}$Institute of Astronomy and Astrophysics, Academia Sinica, 
P.O. Box 23-141, Taipei 10617, Taiwan\\
$^{6}$Space Telescope Science Institute, Baltimore, MD, USA\\
$^7$Dipartimento di Fisica e Astronomia, Universit\`a di Bologna, Viale Berti Pichat 6/2, 40127 Bologna, Italia\\
$^8$Dipartimento di Fisica, Sapienza Universit\`a di Roma and ICRA, p.le A. Moro 2, I-00185 Rome, Italy\\
$^{9}$Dipartimento di Fisica, Universit\`a di Napoli Federico II, Via Cinthia, I-80126 Napoli, Italy\\
$^{10}$INFN Sez. di Napoli, Compl. Univ. Monte S. Angelo, Via Cinthia, I-80126 Napoli, Italy
}

\date{Draft version \today}  
  
\pagerange{\pageref{firstpage}--\pageref{lastpage}}  
  
\begin{document}  
\maketitle  
 
\label{firstpage}  
%%%%%%%%%%%%%%%%%%%%%%%%%%%%%%%%%%%%%%%%%%%%%%%%%%%%%%%%%%%%%%%%%%%%%%%%%%%%%%%  
\begin{abstract}  
Hubble Space Telescope imaging of the galaxy cluster Abell 1689 has revealed an 
exceptional number of strongly lensed multiply-imaged galaxies, including high-redshift 
candidates.  Previous studies have used this data to obtain the most detailed dark matter 
reconstructions of any galaxy cluster to date, resolving substructures ~25 kpc across.  
We examine Abell 1689 (hereafter, A1689) non-parametrically, combining strongly lensed images
and weak distortions from wider field Subaru imaging, and we incorporate 
member galaxies to improve the lens solution. Strongly lensed galaxies are often locally affected
by member galaxies, however, these perturbations cannot be recovered
in grid based reconstructions because the lensing information is too
sparse to resolve member galaxies. By adding luminosity-scaled member
galaxy deflections to our smooth grid we can derive meaningful
solutions with sufficient accuracy to permit the identification of our own strongly lensed
images, so our model becomes self consistent. We identify 11 new multiply
lensed system candidates and clarify previously ambiguous cases, in
the deepest optical and NIR data to date from Hubble and Subaru. Our
improved spatial resolution brings up new features not seen when the
weak and strong lensing effects are used separately, including clumps
and filamentary dark matter around the main halo. Our treatment means
we can obtain an objective mass ratio between the cluster and galaxy
components, for examining the extent of tidal stripping of the
luminous member galaxies. We find a typical mass-to-light ratios of $M/L_B = 21 \pm 14$ inside 
the $r<1$ arcminute region that drops to $M/L_B = 17 \pm 8$ inside the $r<40$ arcsecond region. 
Our model independence means we can objectively evaluate the competitiveness of stacking cluster lenses
for defining the geometric lensing-distance-redshift relation in a
model independent way.
  
\end{abstract}  
%%%%%%%%%%%%%%%%%%%%%%%%%%%%%%%%%%%%%%%%%%%%%%%%%%%%%%%%%%%%%%%%%%%%%%%%%%%%%%%  
%%%%%%%%%%%%%%%%%%%%%%%%%%%%%%%%%%%%%%%%%%%%%%%%%%%%%%%%%%%%%%%%%%%%%%%%%%%%%%%  

%%%%%%%%%%%%%%%%%%%%%%%%%%%%%%%%%%%%%%%%%%%%%%%%%%%  
\section{Introduction}\label{sect_intro}  
%%%%%%%%%%%%%%%%%%%%%%%%%%%%%%%%%%%%%%%%%%%%%%%%%%%  

\begin{figure*}  
   \includegraphics[width=8cm]{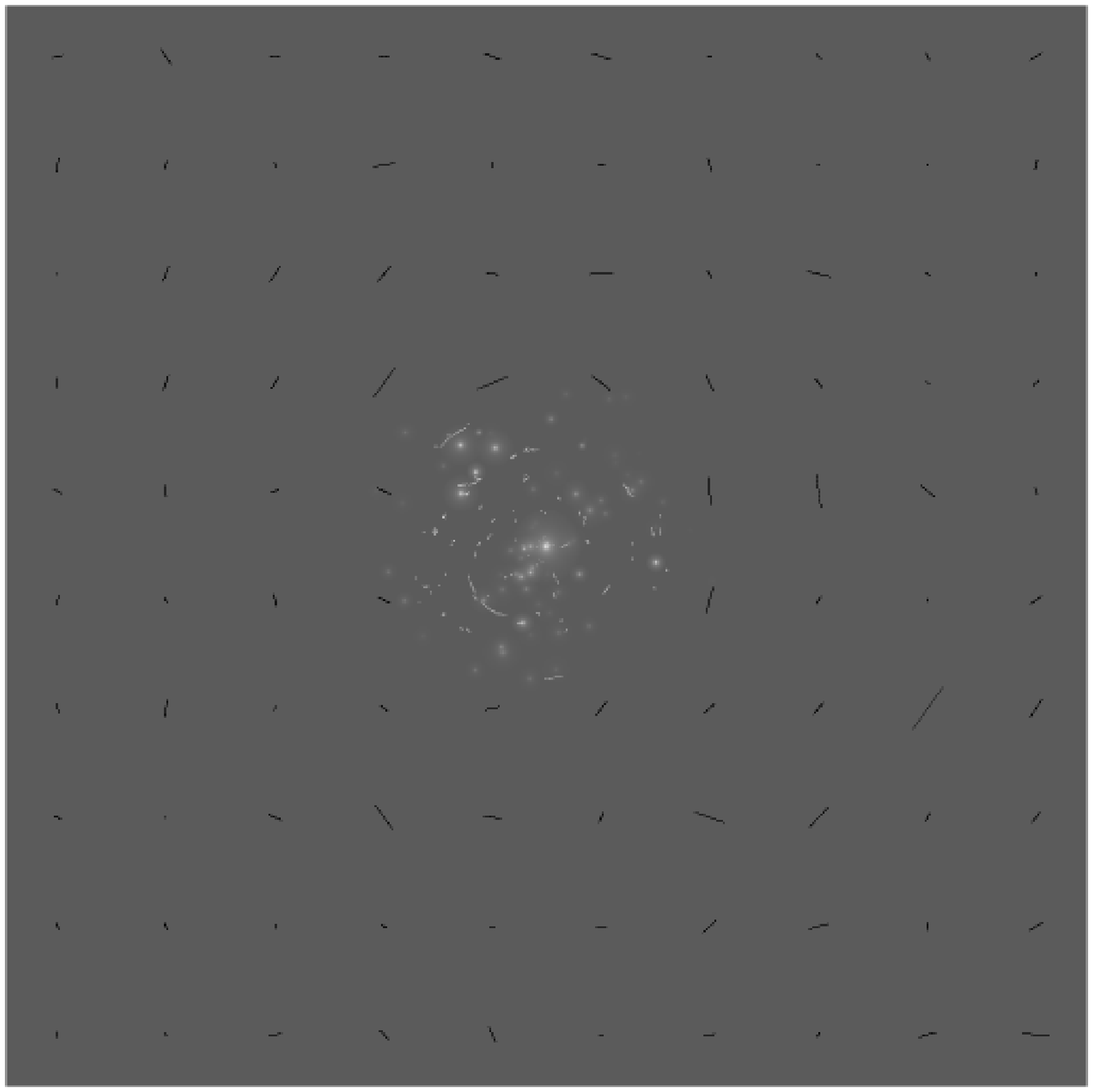} 
   \includegraphics[width=8cm]{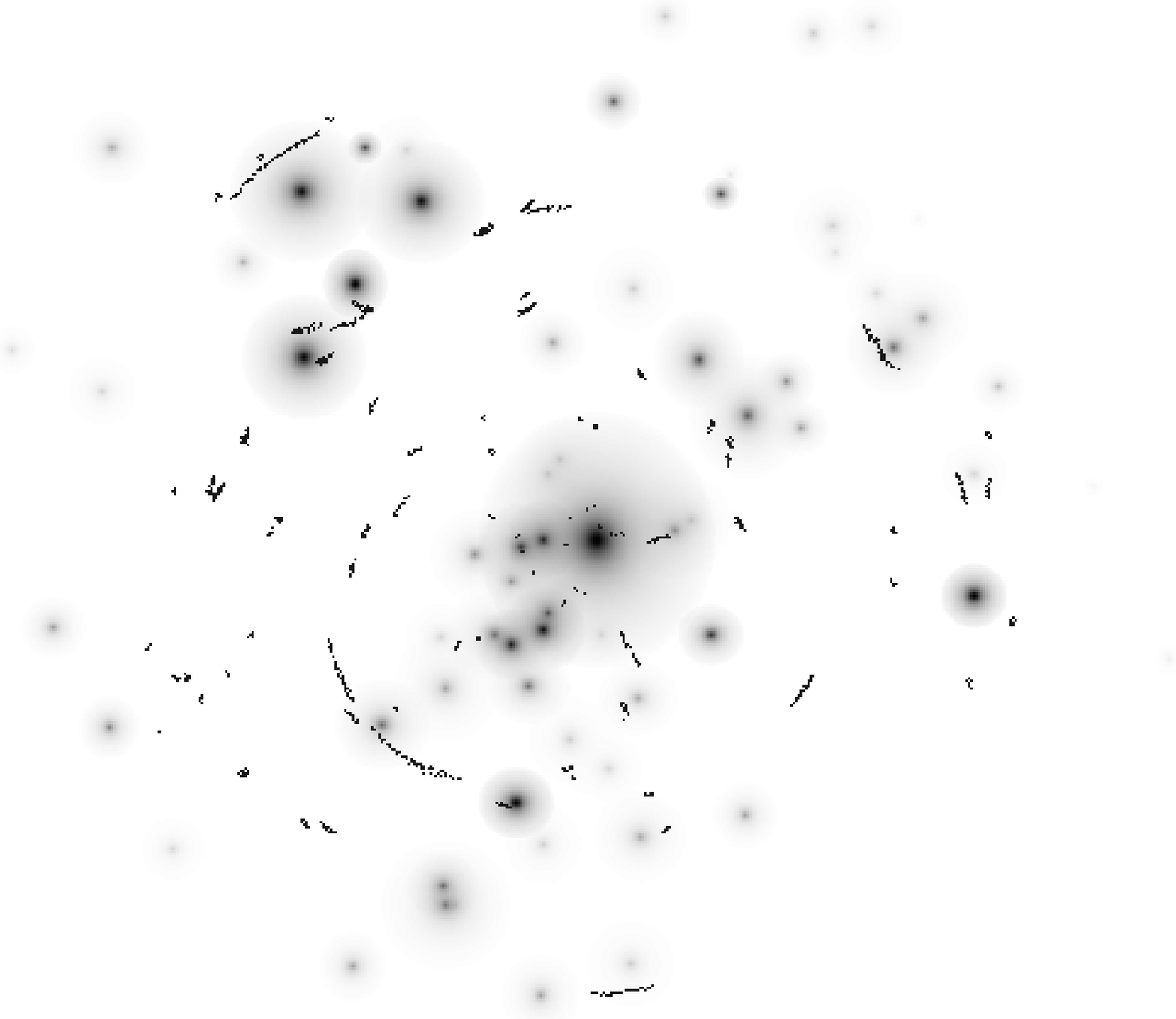}

   \caption{Data set used for the reconstruction. The left panel shows the entire field of view of 
            $10\arcmin$ where the strong lensing arcs are represented in 
            white and the reduced shear measurements are represented in black. 
            The position of the galaxies and their shapes are represented in light grey. 
            The right panel shows the central $3.33\arcmin$ 
            region with the arcs and galaxies in non-linear scale to better show the extent of the galaxy 
            haloes in our model.}  
    \label{fig_data}  
\end{figure*}  

 A fuller exploration of non-parametric cluster lensing is
 increasingly motivated by new dedicated deep Hubble imaging surveys,
 with the aim of examining dark matter structures in the least biased
 way. Multiple sets of lensed images are now typically identified in
 deep, high resolution images of any cosmologically distant cluster
 imaged with Hubble, allowing systematic exploration of the cluster
 dark matter and discovery of the most distant galaxies
 \citep{Broadhurst05a,Clowe2006,Coe2010,Coe2012,Coe2013,Zitrin2009,Zitrin2010,Zitrin2011,
 Zheng2012}.  In practice, secure identification of multiple images
 need the guidance of a reasonably accurate lens model as even the
 counter images of large arcs are typically hard to find given the
 complexities in the central mass distribution of clusters, so that
 images for a given source are far from symmetrically located.

 Furthermore, the uncertain redshifts of faint lensed images means
 that even when a reliable mass model can be built, counter images 
 are predicted to fall on long, largely radial loci or may fail to be
 generated at all if the unknown source distance is 
 insufficient. This means that there are often several contending counter images
 unless internal colors and morphology are sufficiently
 distinctive. Photometric redshifts, if unambiguous, are very helpful
 in limiting the selection of counter images that are too faint
 for spectroscopy. The 16 overlapping broad bands
 of the CLASH program covering the UV to the NIR, maximises the photometric
 redshift accuracy possible with Hubble and have 
 provided reliable examples of the most distant galaxies known, as in
 the case of the $z\simeq 11$ candidate lensed by MACS0647 \citep{Coe2013},
 where multiple images are identified both photometrically and geometrically.

\begin{figure*}  
%   \centerline{ \includegraphics[width=16cm]{figs/A1689_Mean_SNR_Sigma.ps} }                                  
   \includegraphics[width=5.8cm]{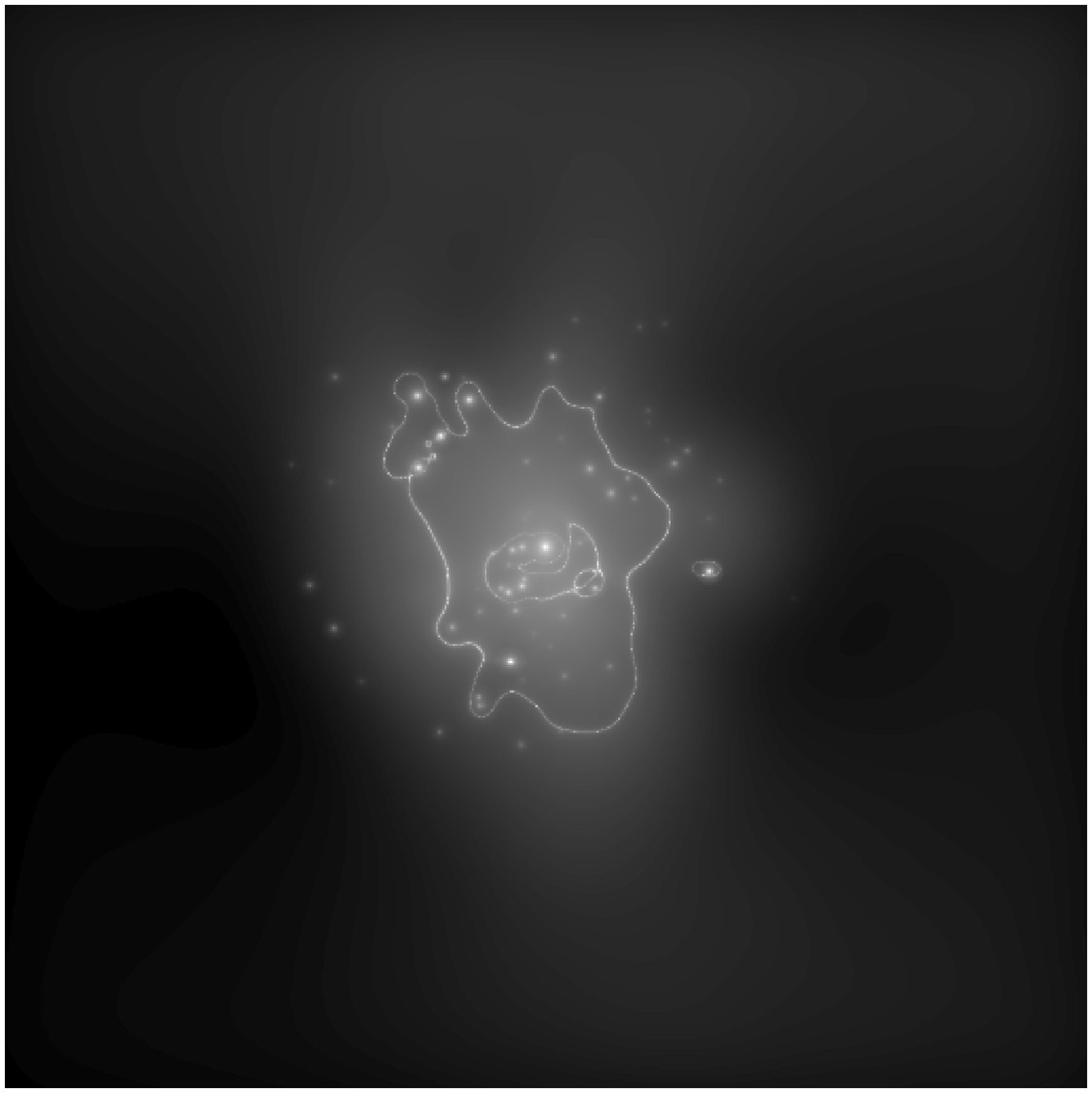}   
   \includegraphics[width=5.8cm]{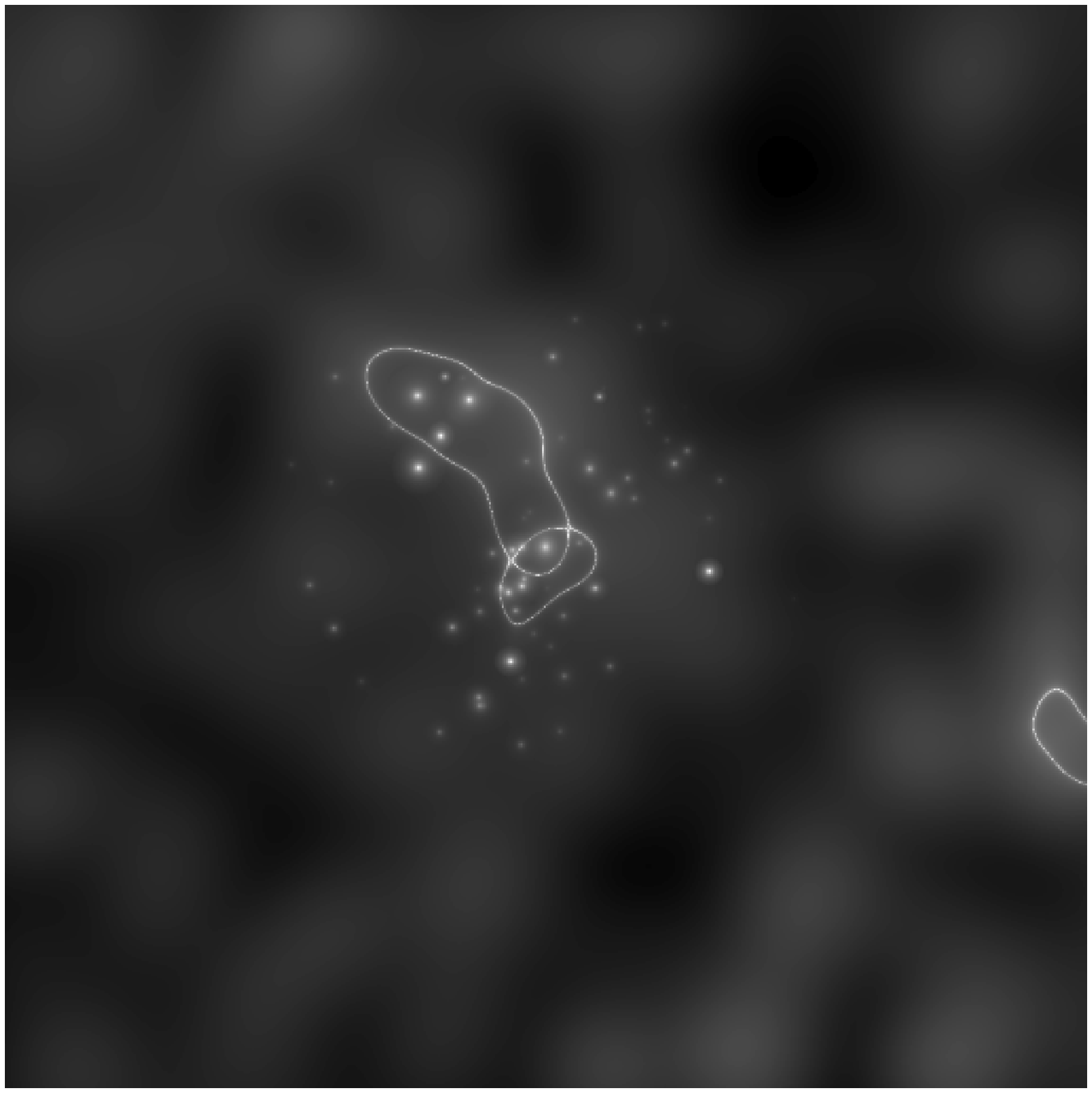}                                   
   \includegraphics[width=5.8cm]{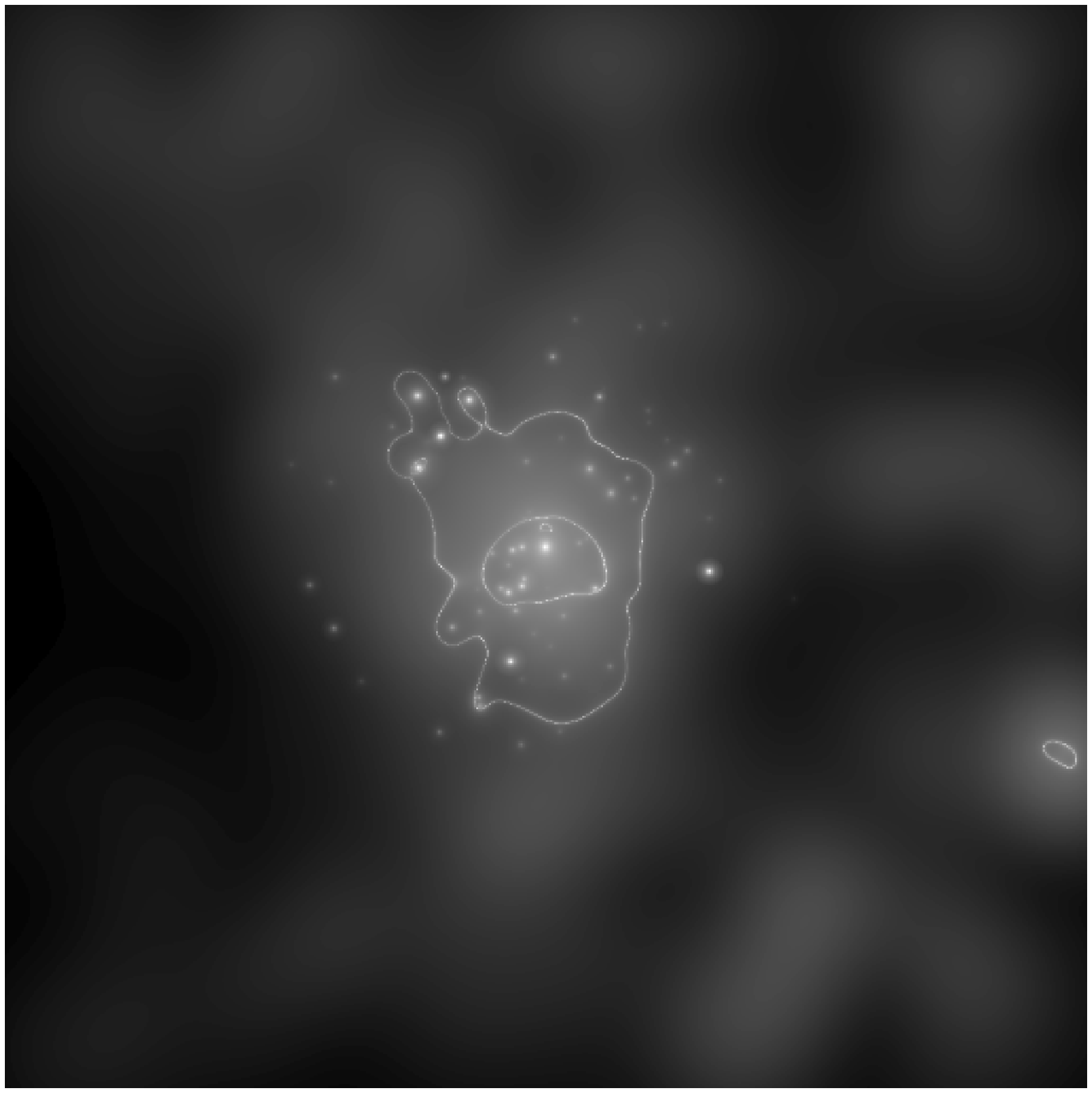}                                   
   \caption{Solutions (and associated critical curves for a source at redshift z=2) for the cases 
            where only the SL data is used in the reconstruction (left), only the WL data is used in the reconstruction 
            (center) and both, the SL and WL data are used in the reconstruction (right). In all cases only the central 
            6.66x6.66 arcmin$^2$ region is shown. 
           }  
   \label{fig_SL_WL_SLWL_CritCurves}  
\end{figure*}  

 To date, the galaxy cluster A1689 remains the best studied cosmic lens with
 hundreds of magnified images in the central region visible in deep
 Hubble images \citep{Broadhurst05a}. Over a hundred of these arcs have
 been matched to their corresponding background galaxies by several
 authors and their redshifts estimated \citep{Broadhurst05a,Halkola2006,Limousin2007,Coe2010}, including several of
 the brightest highest redshift galaxies known, extending to $z\simeq
 7.6$ \citep{Frye2002, Bradley2008}. The relaxed appearance of this cluster and the relatively undisturbed
 optical and X-ray morphology has made this a preferred target for
 constraining the equilibrium mass profile by several independent means \citep{Lemze2008,Sereno2013}.

 Several studies have used these arcs to reconstruct the mass
 distribution using the strong lensing data alone
 \citep{Broadhurst05a,Diego05b,Halkola2006,Jullo2009,Coe2010} and
 in combination with weak lensing (or WL hereafter) measurements
 \citep{Broadhurst05b,Limousin2007,Umetsu2008} including the use of background
 red galaxies whose surface density is depleted by lens magnification
 and independent observationally from weak lensing shear. 
 A1689 has been studied also using higher order derivatives of the lensing potential, like 
 the flexion (see \citep{Leonard2011} for a recent analysis) 
 The mass profile of A1689 was shown to be very well fitted by the standard 
 NFW profile describing the equilibrium mass distribution expected
 for collisionless, cold dark matter (CDM) \citep{Broadhurst05a,Broadhurst05b}
 but with a concentration that is surprisingly high. Triaxiality 
 of the mass distribution has been explored as a means
 of boosting projected concentrations, and certainly may be expected
 to be partially responsible \citep{Oguri2005, Broadhurst2008,Sereno2013}.

 A1689 has been subsequently followed at other wavelengths, allowing
 lensing to be combined with SZ and X-ray data \citep{Sereno2013}
 and also with the dynamics of member galaxy motions via the Jeans
 equation and via velocity caustics \citep{Lemze2009}. 
 Multiwavelength science opens the door to new exciting studies 
 since it is no longer sufficient to model the mass or gas separately
 but instead both have to be integrated in the same model in order to
 explain the observations. 
 Previous work on A1689 combines HST and {\it Chandra} and reveals some
 tension between hydrostatic+lensing reconstruction and other
 observations  \citep{Zekser2006,Leonard2007,Umetsu2008,Cain2011,
 Sereno2011,Lemze2008}.  
 \cite{Peng2009} finds a discrepant hydrostatic mass
 based on X-ray data from {\it Chandra} but \cite{Riemer2009} find
 that excluding substructure alleviates or even eliminates the
 discrepancy.
 Lensing data combined with X-ray and SZ
 data have the potential to reveal information not only about the {\it
 invisible} dark matter distribution but also about the physical
 phenomena taking place in the cluster that have to bring the gas
 pressure and dark-matter-driven gravity to a quasi-equilibrium
 state. The new Frontier Fields program\footnote{http://www.stsci.edu/hst/campaigns/frontier-fields/} 
 is now underway to provide the deepest Hubble data ever recorded for massive lensing
 clusters, further motivating our assumption-free modeling. 
%We aim to provide model independent magnification maps of these clusters 
%so the intrinsic properties of the distant galaxies anticipated in this 
%program can be corrected for lens magnification. 

Despite exhausting lensing studies of A1689, many arcs still remain
unmatched for this cluster. Certainly, many missing counterimages of
highly magnified images are too faint to be useful or remain
undetected. Others are of too low contrast to be detected within
the light of luminous cluster members. Inaccuracy of mass models used
to reconstruct the mass distribution is another issue, given the
significant variation between published solutions. In \cite{Ponente2011},
the authors show how very erroneous mass distributions can still
reproduce lensing data to high accuracy.  In other words, being able
to reproduce the observed arcs is no guarantee that the reconstructed
mass distribution is right. The most detailed analysis made using
A1689 strong lensing data, and that have been able to match tens of
arcs, have relied either on the distribution of member galaxies to
guide the models or on parametric models with the inclusion of many
parameters depending on the number of substructures adopted.

However, the persistent resistance of some obvious bright
lensed images to be matched with other images (when the same models
predict another bright counterpart for those lensed images) suggests
that these models are still missing fundamental pieces that allows to
solve the puzzle. One of the limitations of strong lensing data is
that it quickly becomes insensitive to the mass distribution
beyond the Einstein radius, especially if the distribution of matter
around the center is more or less spherical. Clumps with a significant
amount of matter that lie just beyond the Einstein radius might go
unnoticed with these parametric models as the model contains enough
parameters within the Einstein radius to easily fit the data. Attempts
have been made to constrain the matter distribution beyond the
Einstein radius of A1689 by combining the weak lensing data with strong lensing
data but in all cases (to the best of our knowledge) these {\it joint}
analysis have been made a posteriori where either the density profiles
are combined to extract a single density profile \citep{Umetsu2008}
or the SL solution is tested against the WL data for
consistency \citep{Limousin2007}.

In this paper we revisit this cluster to obtain a truly joint solution
combining in a single inversion (i.e. not {\it a posteriori}) the SL
and WL data. By doing this, our 2-dimensional model of the mass
distribution has to account simultaneously for the multiple lensed
systems observed in A1689 and for the shear measurements that extend
well beyond the Einstein radius.

\begin{figure}  
   \includegraphics[width=8cm]{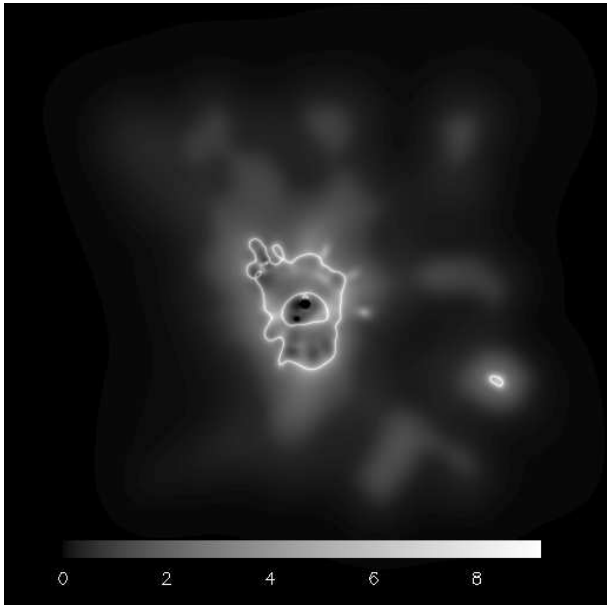} 
   \caption{Magnification map in the entire field of view of 10x10 arcmin$^2$ for the SL+WL case. 
            The colors are in log-scale to increase contrast. 
            The two small black regions in the centre correspond to magnifications less than 1 
            (that have been set to 0 in the log-scale for contrast purposes). 
            The drop in the outer region (buffer zone) is a systematic effect due to the larger cell 
            size in the grid. The features close to the 
            phase transition region between the two grid resolutions are not always to be trusted.}
    \label{fig_magnification}  
\end{figure}  

%%%%%%%%%%%%%%%%%%%%%%%%%%%%%%%%%%%%%%%%%%%%%%%%
\section{ACS data}\label{sect_ACS_data}
%%%%%%%%%%%%%%%%%%%%%%%%%%%%%%%%%%%%%%%%%%%%%%%%
In this paper we used public imaging data obtained from the ACS (filters: F450W and F814W) 
and the WFC3 (filter F125W), retrieved from the Mikulski Archive for Space Telescope (MAST). 
The data come from two different programs. The F814W (ACS) and F125W (WFC3) data were obtained 
within the HST program 11718 (PI Blakeslee, Cycle 17), from May 29 to July 8 of 2010, 
while the ACS F475W images were obtained within the program 9289 (PI Fors, Cycle 11) on June 16 2002.
The total exposure time is 9500~s, 75172~s and 14367~s in the F450W, F814W and F125W filters,
respectively. The F814W dataset has been independently reduced and used by \cite{Alamo2013}
to study the intracluster population of globular clusters.
The data reduction of the optical data consisted in two main steps, based mostly on  
multidrizzle \citep{Koekemoer2002}\footnote{{\tt multidrizzle} is a software tool developed 
by the Science Software Branch at the STSCI and it is appositely designed for to the combination of dithered
images and rejection of cosmic rays.}. 
First, we combined the images obtained in each run and optimized the
image sampling.  Then, we performed  cosmic-rays rejection and aligned 
images\footnote{This step was performed by using the IRAF-geomap package}.
Final mosaic has pixel scale 0.05 arcsec.
We combined these three bands images to produce the color image of the new candidate lenses shown in the Appendix.

%%%%%%%%%%%%%%%%%%%%%%%%%%%%%%%%%%%%%%%%%%%%%%%%
\section{Lensing data on A1689}\label{sect_data}
%%%%%%%%%%%%%%%%%%%%%%%%%%%%%%%%%%%%%%%%%%%%%%%%
A compilation of systems found in the literature is shown in table 2
in the appendix. They are obtained basically from 3 sources,
\cite{Broadhurst05a}, \cite{Limousin2007} and \cite{Coe2010}.  
Many of these systems are also listed in \cite{Halkola2006}. 
Table 2 is  built after cross-correlating the original tables in the 
references above to avoid repetitions. Systems that were listed originally in \cite{Broadhurst05a} are 
referred as B05 in table2. Systems that appear in both  \cite{Limousin2007} and \cite{Coe2010} or just in 
\cite{Coe2010} are listed as C10.  Systems that appear only in \cite{Limousin2007} are listed as L07 and the 
new system candidates presented in this paper are listed as D14 in table 2. 
Some of the original systems in \cite{Broadhurst05a} were rearranged or updated with additional arclets by 
other authors in later papers after comparison with alternative mass models. 
In the present paper we will rely on the original selection of \cite{Broadhurst05a} 
after excluding some dubious systems but will explore also the solutions obtained 
after incorporating the alternative systems published in the literature. The exclusion (or 
re-arrangement) of some systems listed in table 2 is made after a new visual color and morphology 
comparison of the system members based on the new and deep ACS images. For instance, the confusion between 
systems 10 and 12 can be resolved by the presence of a pinkish core in system 10 not present in system 12.   
Different authors (\citep{Limousin2007} and \citep{Coe2010}) have suggested alternative rearrangements for some of 
these systems. Although we do not consider all the alternative possibilities in this paper, they might  
be perfectly valid as well. In fact, as will be shown later, some of the new counterimages discovered 
by other authors (like in system 12) will be naturally predicted/confirmed by our model, and hence, 
fully consistent with it.   
If the system was listed in \cite{Broadhurst05a} we maintain the system identification from 
\cite{Broadhurst05a}. Alternative identifications of the central counterimages of some systems 
have been used in \cite{Limousin2007} and \cite{Coe2010}. Some of these alternative identifications 
are also reflected in table 2.  
Although not explicitly mentioned in the table 2, the last system in
\cite{Halkola2006} corresponds (at least partially) to our system 58
in table 2.  The last 11 systems of table 2 (denoted with D14 in the REF
column of table 2) should be treated just as mere candidates since they are
obtained after identifying new system candidates using our solution
discussed below and without a proper photometric redshift estimation. 
The new candidate systems are shown in figures
\ref{fig_StampsI} and \ref{fig_StampsII} in the appendix. The stamps are extracted 
from a RGB composite image of 3 Hubble filters (F475w, F814w and the near infrared F125w).    
Some of the positions listed in \cite{Coe2010} where also incorrectly
translated into the tex file in the original paper. 
These positions have been corrected in table 2. Table 2 contains a total of 50 systems.
{\footnote{A full set of stamps from this table can be found at 
http://max.ifca.unican.es/diego/FigsA1689/}}  
Systems 25 and 32 contain multiple candidates
for some of the secondary images and will not be used in our
analysis. Out of the 50 systems listed in table 2, we use only a
reliable subset of 26 systems corresponding to systems 1 through 30 in
\cite{Broadhurst05a} but excluding suspicious systems 20, 26, 27 (in addition to system 25 mentioned above) 
System 10 and 12 have been re-matched after examination of new IR data that exhibit 
distinctive colors. 

Our WL data are derived from \cite{Umetsu2008} based on Subaru $Vi'$ imaging observations, and we refer the reader to 
that paper for a detailed description of their observations and analysis.  
In this work, we use two-dimensional reduced-shear data on a regular grid of $10 \times 10$ independent 
grid points with 
$1\arcmin$ spacing, covering the central $10 \times 10$ arcmin$^2$ region.  We exclude from our analysis the innermost 
four pixels overlapping with the critical-lensing regime, so that our WL data set consists of $100-4=96$ reduced-shear 
data points. Since the fields of view and orientations of the HST and Subaru data sets are different (in their native 
form), we rotate and re-centre the SL data to match the centroid and orientation of the WL data. 
Figure 1 (left) shows the two data sets used in this paper as well as the distribution (and morphology) of member 
galaxies that will be used to build the fiducial deflection field (see the next section). The right panel shows the 
central $3.3\times 3.3$ arcmin$^2$ region (WL data not shown).

%%%%%%%%%%%%%%%%%%%%%%%%%%%%%%%%%%%%%%%%%%%%%%%%%%
\section{Reconstruction method}\label{sect_method}
%%%%%%%%%%%%%%%%%%%%%%%%%%%%%%%%%%%%%%%%%%%%%%%%%%
\begin{figure}  
   \includegraphics[width=9cm]{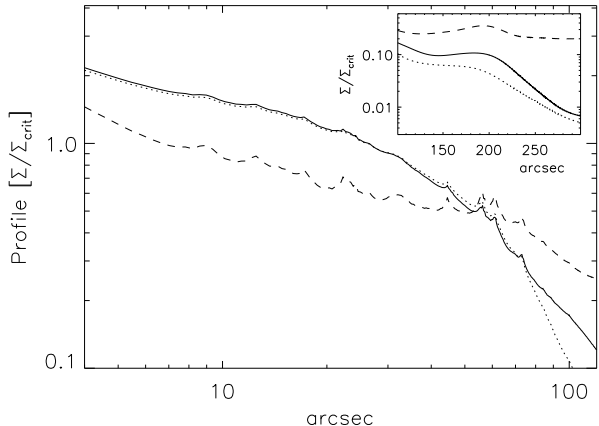} 
   \caption{Profiles of the solutions obtained with SL data only (dotted), WL data only (dashed) and the combination 
            SL+WL (solid). The smaller plot shows the same profiles but in linear scale and beyond $100\arcsec$. 
            The units in the axis are the same as the larger plot. 
            The systematic bump at $200\arcsec$ (or $3.33\arcmin$) 
            coincides with the position of the transition phase (for the grid) between the $6.66\arcmin$ region 
            and the outer buffer zone}
    \label{fig_Profiles_SL_WL_SLWL}  
\end{figure}  
We use the improved method, WSLAP+, to combine the weak and strong
lensing data and perform the mass reconstruction. The reader can find
the details of the method in our previous papers
\citep{Diego05a,Diego05b,Diego2007,Ponente2011,Sendra2014}. Here we
give a brief summary of the most essential elements. \\

Given the standard lens equation, 
\begin{equation} \beta = \theta -
\alpha(\theta,\Sigma(\theta)), 
\label{eq_lens} 
\end{equation} 
where $\theta$ is the observed position of the source, $\alpha$ is the
deflection angle, $\Sigma(\theta)$ is the surface mass density of the
cluster at the position $\theta$, and $\beta$ is the position of
the background source.  Both the strong lensing and weak lensing
observables can be expressed in terms of derivatives of the lensing
potential. 
\begin{equation}
\label{2-dim_potential} 
\psi(\theta) = \frac{4 G D_{l}D_{ls}}{c^2 D_{s}} \int d^2\theta'
\Sigma(\theta')ln(|\theta - \theta'|), \label{eq_psi} 
\end{equation}

where $D_l$, $D_{ls}$ and $D_s$ are the
angular diameter distances to the lens, from the lens to the source
and from the observer to the source, respectively. The unknowns of the lensing
problem are in general the surface mass density and the positions of
the background sources. As shown in \cite{Diego2007}, the
weak and strong lensing problem can be expressed as a system of linear
equations that can be represented in a compact form, 
\begin{equation}
\Theta = \Gamma X, 
\label{eq_lens_system} 
\end{equation} 
where the measured strong and weak lensing observables are contained in the
array $\Theta$ of dimension $N_{\Theta }=2N_{SL} + 2N_{WL}$, the
unknown surface mass density and source positions are in the array $X$
of dimension $N_X=N_c + N_g + 2N_s$ and the matrix $\Gamma$ is known
(for a given grid configuration and fiducial galaxy deflection field, 
see below) and has dimension $N_{\Theta }\times N_X$.  $N_{SL}$ is the number
of strong lensing observables (each one contributing with two constraints,
$x$, and $y$) $N_{WL}$ is the number of weak lensing observables
(each one contributing with two constraints, $\gamma_1$, and $\gamma_2$),
$N_c$ is the number of grid points (or cells) that we use to divide
the field of view.  $N_g$ is the number of deflection fields (from
cluster members) that we consider.  $N_s$ is the number of background
sources (each contributes with two unknowns, $\beta_x$, and $\beta_y$,
see \cite{Sendra2014} for details). The solution is found after
minimizing a quadratic function that estimates the solution of the
system of equations \ref{eq_lens_system}.  For this minimization we
use a quadratic algorithm which is optimized for solutions with the
constraint that the solution, $X$, must be positive. This is
particularly important since by imposing this constraint we avoid the
unphysical situation where the masses associated to the galaxies are
negative (that could otherwise provide a reasonable solution, from the
formal mathematical point of view, to the system of linear equations
\ref{eq_lens_system}). Imposing the constrain $X>0$ also helps in regularizing 
the solution as it avoids large negative and positive contiguous fluctuations. 

\begin{figure}  
   \includegraphics[width=4cm]{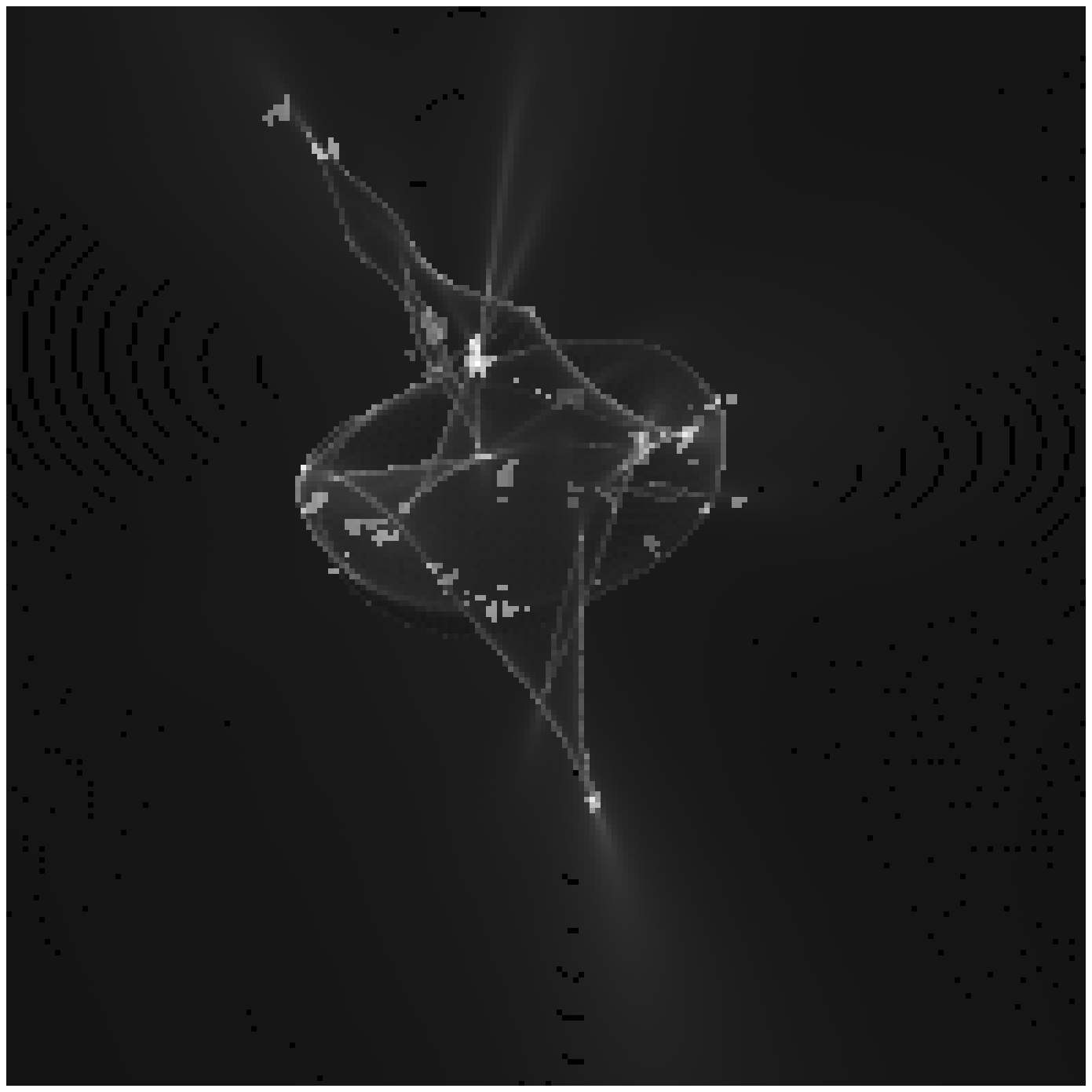}  
   \includegraphics[width=4cm]{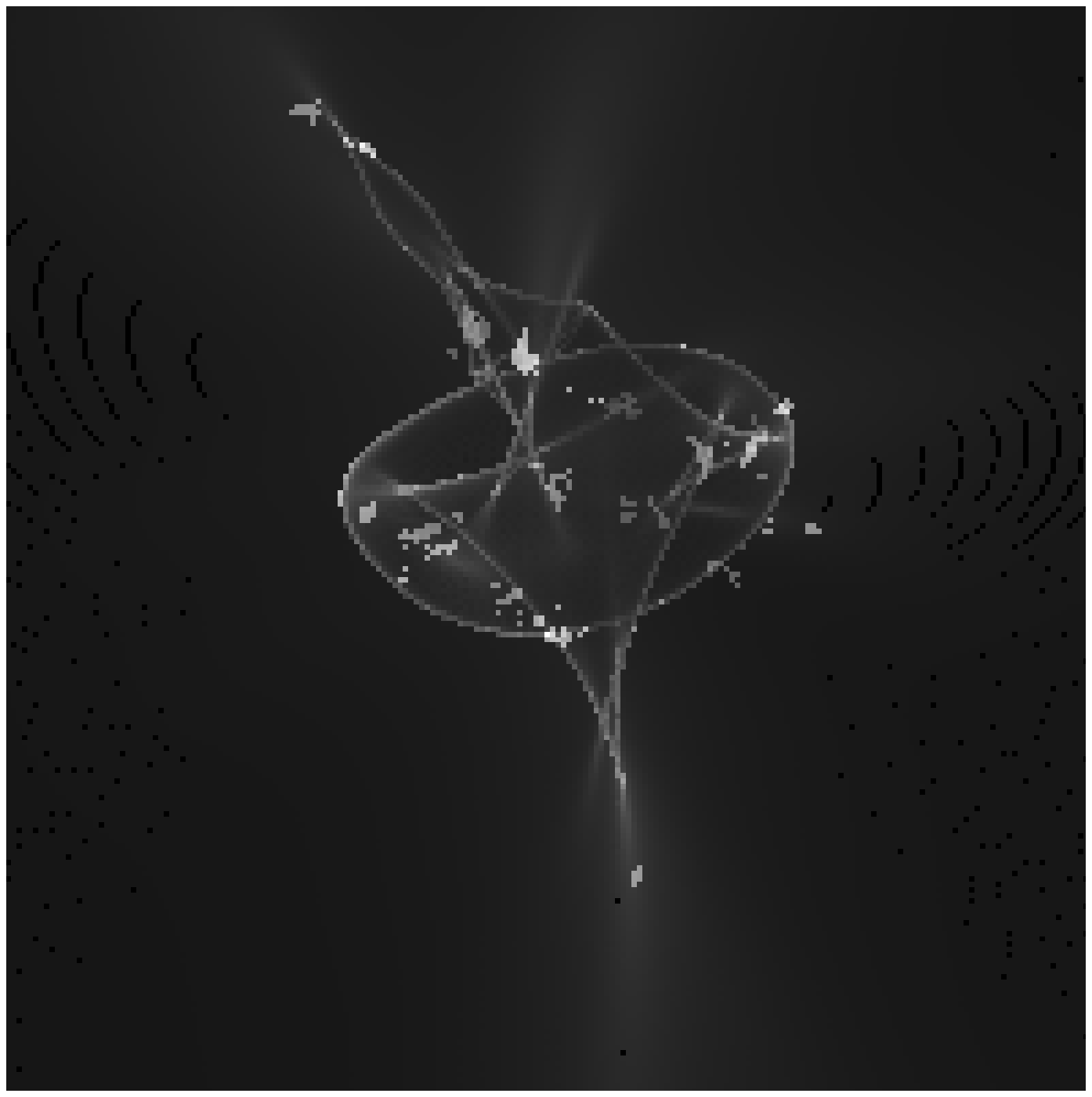}  
   \caption{Caustics for the SL only (left) and SL+WL case (right) compared with the reconstructed sources. 
            Each source is represented with a different color index from 1 (darkest grey) to 26 (lightest grey). 
            The field of view corresponds to $1.3\arcmin$ across. The centre is the same in both cases and 
            corresponds to the centre of the original $10\arcmin$ field of view. }
    \label{fig_Caustics_Beta_SLWL}  
\end{figure}

Earlier work has shown how the addition of the small deflection fields from member galaxies 
can help improve the mass  determination when enough constraints are available (see for instance 
\cite{Kassiola1992}, \cite{Kneib1996}). In our previous paper \citep{Sendra2014} we quantified via
simulations how the addition of deflections from all the main member
galaxies helps improve the mass reconstruction with respect to our previous standard 
non-parametric method. For our study we select the 73 brightest elliptical galaxies (from the red sequence) 
in the cluster central region and associate to them a mass according to their luminosity. 
We assume the fiducial deflection field comprising these member galaxies just scales 
by a fixed luminosity--mass ratio.  
Later, the fitting procedure determines this proportionality constant that allows 
for the best reproduction of the data.  
In \cite{Sendra2014} we used one deflection field to model all the
galaxies in the cluster. In the case of A1689 we go a step further and
we use two deflection fields (i.e $N_g=2$, see definition of $N_g$
above).  The first one is associated to the central type-cD galaxy and
the second one contains the deflection field from the remaining dominant
galaxies in the cluster. Each deflection field contributes in our
model as one free parameter (its amplitude with respect to the
fiducial amplitude). In principle one could incorporate an independent
deflection field for each one of the member galaxies but caution has to be taken to
maintain as much as possible the orthogonality between the grid cells
and the individual deflection fields. However, this is an interesting alternative 
that will be explored further in the future.  Settling for two deflection
fields may be regarded as a fair compromise between the overly simple
assumption that all galaxies in the cluster have individual halos with
masses that trace light following the same luminosity-mass relation
and a potentially more realistic but also unnecessarily complex
assumption that each galaxy has a different luminosity-mass ratio.  We
make an exception for the central cD galaxy because of its distinctive
shallow luminosity profile and the separate origin that may be implied
by the anomalously large numbers of globular clusters for this object
and cD galaxies in general (see \cite{Alamo2013}). 
All the galaxies used in our fiducial model are shown in figure
\ref{fig_data} where we use a non-linear color scale to better show
the extent and shapes of the individual haloes in our fiducial
model. As in our previous paper \citep{Sendra2014}, we consider
truncated NFW profiles to construct our fiducial model. 

%%%%%%%%%%%%%%%%%%%%%%%%%%%%%%%%%%%%%%%%%%
\section{Results}\label{sect_results}
%%%%%%%%%%%%%%%%%%%%%%%%%%%%%%%%%%%%%%%%%%
When combining the WL and SL data sets, due to the large field of
view (10x10 arcminutes$^2$ sampled with a total of $1536^2$ pixels), 
and in order to maximize the resolution of the grid in the
region covering the SL part of the data, we use a two resolution grid
where the central 6.66x6.66 arcmin$^2$ is sampled with cells of 24x24
pixels and the remaining area is sampled with 64x64 pixels cells.  
The use of a multiresolution grid introduces a bias in the reconstruction (see
discussion below) in the transition region between the two
resolutions. The outer region is used as a buffer zone that, however,
still contributes to the WL constraints in the transition phase
between the two regions. Hence, we don't use the results from the outer
region in our conclusions but instead we will restrict ourselves to
the central 6.66x6.66 arcmin$^2$ region. However, we should note that
even within the central region caution has to be taken when
interpreting the results close to the transition phase as some biases
are still present near the border.

\begin{figure}  
\centerline{\includegraphics[width=8cm]{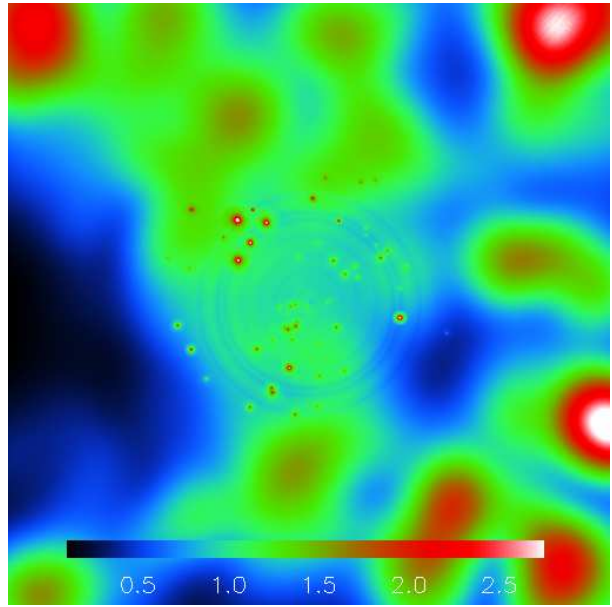}}      
   \caption{ 
            Asymmetric behaviour of the mass density in the central region of 6.66x6.66 arcmin$^2$. 
            The plot shows the ratio between the solution 2D map (SL+WL case) and 
            the corresponding profile, that is, each pixel at a distance $r$ from the centre is divided by the 
            profile at the same $r$. In order to increase contrast the ratio is saturated beyond the value 7.5 
            and we show the square root of this ratio. 
           }  
   \label{fig_Mass2profile_ratio}  
\end{figure}

For comparison purposes, we have performed the reconstruction in three
different cases depending on the data set used. In case (i) we use
only the SL data set, in case (ii) we use only the WL data set, and in
case (iii) we combine the SL and WL data set into the same data
vector. In order to make a direct comparison, we use the same grid for
all three cases although this is not optimal for the SL nor WL
case. In the SL-only case, we would use only a regular grid covering a
smaller field of view (of 3x3 arcmin$^2$) while in the case of the
WL-only case we would also use a regular grid (with poorer resolution) 
but over the entire10x10 arcmin$^2$ field of view. Since our main interest 
is on the solution obtained when the SL and WL data sets are combined, we
maintain the same grid configuration in all three cases. Also, we
start the minimization in the same initial condition to eliminate this
degree of freedom from the solution (different initial conditions are
explored later in the paper). Finally, we use the same number of
iterations (8000 iterations, this number will be discussed later) in
the SL-only and SL+WL cases. For the WL-only case we stop the
minimization before to avoid over-fitting (a large number of iterations
in the WL-only case would produce a solution that is capable of
reproducing the noisy WL estimates so the minimization must be stopped
before this regime is reached). In figure
\ref{fig_SL_WL_SLWL_CritCurves} we present the 2D reconstructed mass
in the central region and the associated critical curves for a source
at redshift 2 for the three cases.  From left to right we show the
cases of the SL-only, WL-only and SL+WL. Note how the critical curve
does not change much in the case SL+WL when compared with the SL-only
case. This is a consequence of the critical curve being much more
sensitive to the very central region (and the SL data). However, when
the WL data is used in combination with the SL data, new interesting
features in the mass distribution emerge even beyond the Einstein
radius. Some of these features appear even more intriguing when
looking at the magnification map (see figure \ref{fig_magnification})
where some filamentary structures and clumps are made more evident.
Both the projected mass and magnification maps are closely connected
and the magnification map in this weaker lensing regime can be used as
an alternative tracer of the mass especially in the range of interest
shown in figure \ref{fig_magnification} where the convergence,
$\kappa<0.5$, and $\kappa \sim \gamma$.  In this regime, and to first
order, the magnification, $\mu$, can be approximated (by Taylor
expansion) as $\mu \approx 1 + 2\kappa$ that shows the clear
connection between magnification and projected mass. Further investigation
of these features demands better quality WL data and will be
the subject of future studies.  The small critical curve around the
clump at the edge of the bottom right quadrant is close to a feature
seen also in \cite{Umetsu2008}. This fact suggests that the feature in
our reconstruction, although it could be affected by its proximity to
the buffer zone, may also be produced by an enhancement in the
magnitude of the WL signal in that area.

A more quantitative comparison of the different solutions is shown in
figure \ref{fig_Profiles_SL_WL_SLWL} where we show the density
profiles of the surface mass density in terms of critical
density. Unless otherwise noted, the critical density,
$\Sigma_{crit}=4.746\times10^{15} M_{\odot}h/Mpc^2$, is computed for
$z_{mean}=1.07$ (as in previous work).  The WL-only case shows the
typical mass-sheet degeneracy which has not been corrected in our
solution. Also, in the WL-only solution, the mass at the centre is
mostly associated with the individual galaxies accounting for the shear
in the vicinity of the Einstein radius, while the grid complements the
central galaxy mass. On the other hand, in the SL-only case, the grid
plays a more central role and accounts for most of the mass. Also, the
SL-only case shows a sharp drop in mass beyond the Einstein radius,
which is expected for this grid based model, due to the lack of
sensitivity of the SL data to the outer regions. When the SL and WL
data sets are combined in the joint reconstruction, the new joint
profile shows a smoother behaviour to larger radius and the solution
compares well with previous estimates of the profile derived from
SL-only and WL-only. Both grid and galaxies play important roles 
in fitting the SL+WL data set.

  In  terms of source reconstruction, figure \ref{fig_Caustics_Beta_SLWL}
 shows the reconstructed sources for the two cases, SL-only (left) and
 SL+WL (right). Both, scale and image's centre are the
 same. In both cases, the solutions obtained with SL-only and SL+WL
 data seem to be able to form $1\arcsec-3\arcsec$ sources that fall
 near well defined caustics.

In figure \ref{fig_Mass2profile_ratio} we show the ratio between the
2D reconstructed mass map and the corresponding profile. A circularly 
symmetric mass distribution should behave as a constant sheet of value
$Ratio=1$. Deviations from this value highlight the asymmetries in
different regions of the cluster. Note how the central region exhibits
a more symmetric structure but around the Einstein radius there are
important deviations from the mean profile by a factor $\approx 4$
above and below the mean density. The largest deviations occur near
the buffer zone, and may be affected by the proximity to this
transition phase. Overall, a left-right global asymmetry (or gradient)
is appreciated across the field of view. A similar asymmetry can also
be found in \cite{Umetsu2008}.  The rings around the centre are due to
the individual galaxies which produce spikes in the profile.

\subsection{Comparison with previous results and analytical models}
\begin{figure} \centerline{
\includegraphics[width=9cm]{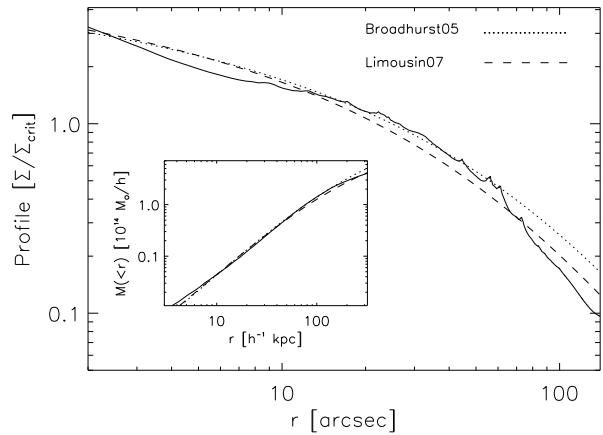}}
   \caption{Comparison of the profile corresponding to the SL+WL solution (after 8000 Iterations) with two NFW models 
            found in the literature, and that fit their corresponding solutions \citep{Broadhurst05b,Limousin2007}}
   \label{fig_integrated_Mass}  
\end{figure}

%%%%%%%%%%%%%%%%%%%%%%%%%%%%%%%%%%%%%%%%%%%%%%%%%%%%%%%%%%%%%%%%%%%%%%
It is important to compare the results obtained with our
non-parametric algorithm with those obtained using fully parametric
methods but a similar data set. Figure \ref{fig_integrated_Mass} shows
our solution compared with two analytical models that fit solutions
obtained by parametric methods. \cite{Broadhurst05b} found that in
detail the NFW profile did not fit well the SL+WL data in a fashion
similar to our case, with a tendency to be too steep in the center ($r<50$kpc).

\begin{figure}  
\includegraphics[width=9cm]{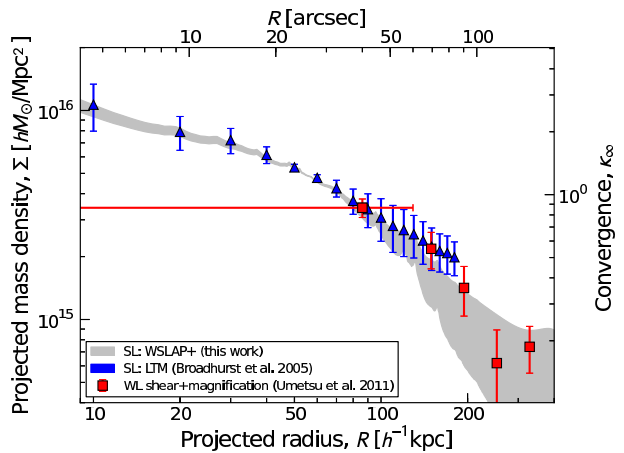}
   \caption{Comparison of the profiles from our solution (grey region from table 1) with previous results from 
            the literature.}
   \label{fig_profiles2}  
\end{figure} 

In figure \ref{fig_profiles2}   we compare the radial mass profile derived from our solution with previously published 
results of A1689 based on different lensing techniques. In the SL regime, our solution overlaps well with the SL 
modeling results of Broadhurst et al. (2005a). Our results are also in good agreement with the model-independent mass 
profile of \cite{Umetsu2011} derived from combined WL shear-and-magnification measurements based on the Subaru data.
When compared with previous work, we find a good agreement between our solution and other solutions, in terms of 
the profile and the location and shape of the radial critical curve. Although the tangential critical curve 
we obtain extends further to the south than previous SL solutions.

\subsection{Mass-to-light ratios}
%%%%%%%%%%%%%%%%%%%%%%%%%%%%%%%%%%%
Since our mass model has the galaxy member component differentiated from the diffuse dark matter halo component 
we can compute light-to-mass ratios at the position of the member galaxies. We compute the luminosity (B-Johnson) 
of the galaxies over the same area covered by our fiducial mass model. 
The mass-to-light ratio oscillates around a typical value of around 20 for most of the galaxies, with a small 
decrease towards the central galaxy. In particular, we find a mass-to-light ratio of $M/L_B = 21 \pm 14$ inside 
the $r<1$ arcminute region that drops to $M/L_B = 17 \pm 8$ inside the $r<40$ arcsecond region. 
Recently,  \cite{Okabe2013} found that at large cluster radii, the mass-to-light ratio of sub-halos in the Coma 
cluster tend to the typical values for clusters (around 200), whereas this ratio decreases towards the 
centre of the cluster to values around $M/L_B \approx 35$ (for $h=0.7$) (see also, \citep{Natarajan2009}).  
The fact that our critical curves present a smoother form when compared to previous estimations (see for 
instance \cite{Broadhurst05a,Halkola2006,Limousin2007,Coe2010}) while the total mass inside the Einstein radius 
is consistent with previous work, suggesting that the masses associated to the individual galaxies in our solution 
are smaller than the corresponding masses derived from alternative methods. The mass-to-light ratio inferred from 
our solution is therefore probably smaller than the one that could be derived from those alternative methods. 

\section{Cosmological implications}\label{sect_cosmo}
%%%%%%%%%%%%%%%%%%%%%%%%%%%%%%%%%%%%
\begin{figure}  
%\centerline{ \includegraphics[width=9cm]{figs/fk.eps} }                     
\centerline{ \includegraphics[width=9cm]{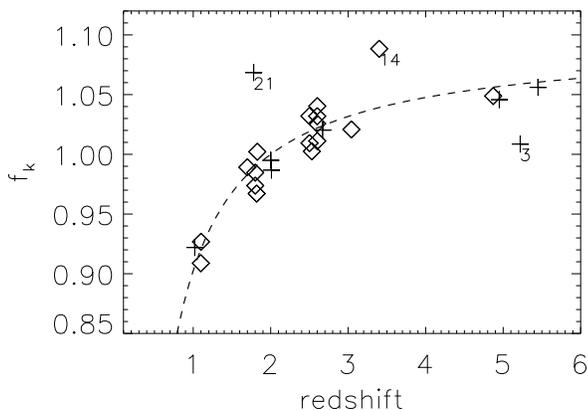} }                     
   \caption{ 
            The dashed line corresponds to the function $f_k$ for a flat model 
            ($\Omega_m = 0.3$, $\Lambda=0.7$) compared with the 
            data (diamonds correspond to systems with spectroscopic redshifts and crosses to systems 
            with photometric redshifts). The main outliers are marked 
            with their corresponding ID. 
           }  
   \label{fig_fk}  
\end{figure}  
\begin{figure*}  
%   \centerline{ \includegraphics[width=16cm]{figs/A1689_Mean_SNR_Sigma.ps} }                                  
   \includegraphics[width=5.8cm]{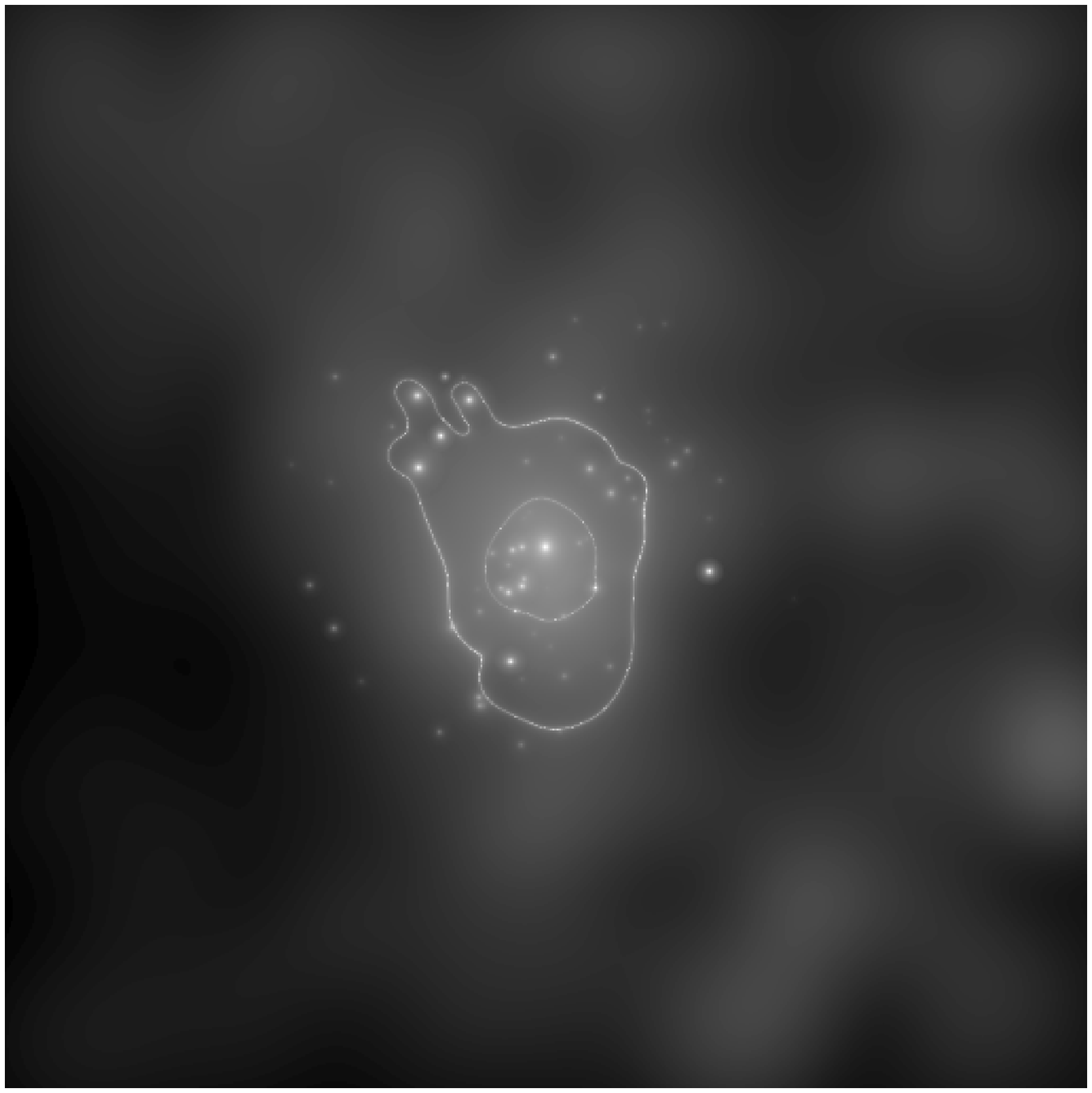}
   \includegraphics[width=5.8cm]{figs/Critical_curve_SLWL_8000IT.ps}
   \includegraphics[width=5.8cm]{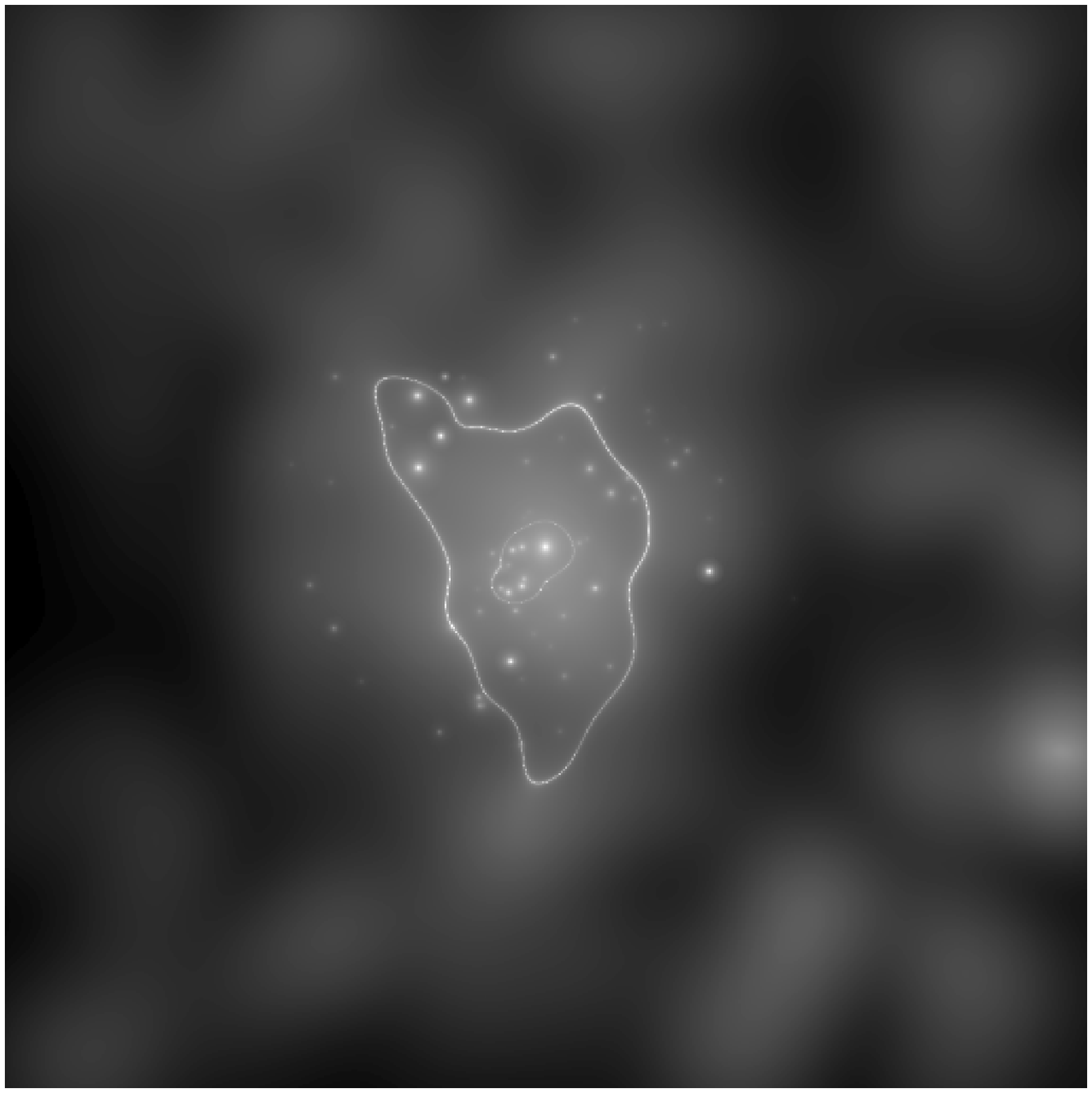}
   \caption{Solutions (and associated critical curves for a source at redshift z=2) for three different iteration numbers. 
            Left corresponds to 5000 iterations, middle to 8000 iterations and right to 15000 iterations. 
            Note how for 8000 iterations both radial and tangential critical curves fall very close to radial and 
            tangential arcs.}  
   \label{fig_SLWL_CritCurves_5000_8000_15000}  
\end{figure*}  
Gravitational lensing in well studied clusters like A1689 can be used to impose constraints 
in the cosmological model (see for instance \citep{Jullo2010}). A test based on the relative 
differences of the deflection angle between pairs of images was applied to A1689 in 
\cite{Broadhurst05a} based on the solution obtained with a parametric model.
Figure \ref{fig_fk} shows the $f_k = D_{ls}(z)D_z(z_s=2)/D_s(z)D_{ls}(z_s=2)$ function as described in
\cite{Broadhurst05a} (see eqs. 7 and 14 in that paper). This function is normalized for convenience to
$z=2$ and the shape is determined by the cosmological model.  
%although it is insensitive to the parameter range now "available" from other
%distance related measurements. 
    Each data point correspond to
    a multiply lensed system (out of our set of 26). Due to the more unprecise 
    reconstruction in the very centre of the cluster, we exclude images that are 
    at a distance of 5 arcseconds or less from the centre. Triangles correspond to the
    spectroscopic redshift systems and cross symbols to those with only
    photometric redshifts.  The dashed line indicates the
    expected behaviour of our data points for a standard cosmological
    model (flat $\Lambda$CDM model with $\Omega_m=0.3$). In the ideal
    scenario where there is no projection effects and we are able to
    reconstruct the deflection field perfectly, the data points would
    lie perfectly along the curve for the correct choice of cosmological
    model. Some scatter is seen here about this expected
    relation. The symmetry of the scatter indicates that the
    deflection field we recover is not noise free and imperfections and
    projection effects along the line of sight unrelated to the
    cluster must at some level limit the accuracy of this
    comparison. However, it is important to notice
    that our result is obtained with just one cluster and is not
    optimized for in terms of the "best" multiply lensed systems.  For
    instance, 3 systems are marked in the above plot that depart more significantly from the
    expected theoretical behaviour. System number 3 corresponds to a
    system with only a photometric redshift (and this redshift could be
    wrong) and besides, two of the images of this system are
    relatively close to each other (this close proximity of course
    enhances the uncertainty in estimating $f_k$ for such systems)
    and the third image is basically
    buried behind one of the large elliptical galaxies (and hence very
    sensitive to the exact mass distribution of this member galaxy at a
    level not incorporated in our member galaxy model. A similar situation 
    is found in system 21, where two radial images are close to the centre (although 
    farther than the 5 arcsecond exclusion radius mentioned above) and one of them is very 
    close to one of the member elliptical galaxies. System 14 corresponds to a system that
    lies well beyond the Einstein radius, where our reconstruction
    is less well constrained. It is obvious from the
    above plot that a more accurate description of the lens (for
    instance through the addition of new spectroscopic systems,
    especially at high redshift) would permit a tighter constraint on
    the cosmological model.  A study based
    on stacking $f_k$ for many lensing clusters should be able to
    provide competitive constraints on the cosmological model based on
    this potentially important independent test, for which some simulations 
    have been explored \citep{Lubini2014}. 

\begin{figure}  
   \includegraphics[width=9cm]{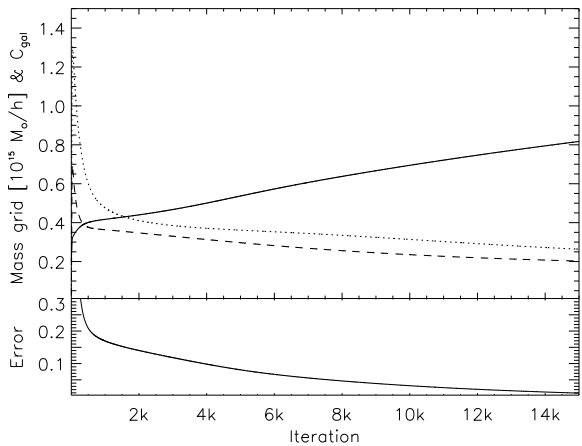}
   \caption{Top panel. Total mass in the grid (solid line) versus iteration number (from 1 to 15000). 
            The dotted and dashed lines show the two coefficients ${\rm C}_1$ and ${\rm C}_2$ respectively. 
            These coefficients multiply the fiducial fields for the type-cD (${\rm C}_2$) galaxy and the 
            remaining galaxies (${\rm C}_1$). The bottom panel shows the value of the function that is being 
            minimized as a function of iteration number 
            (the units have been re-scaled by a constant for clarity purposes)}
    \label{fig_Iter_M_C}  
\end{figure}

\begin{figure}  
   \includegraphics[width=9cm]{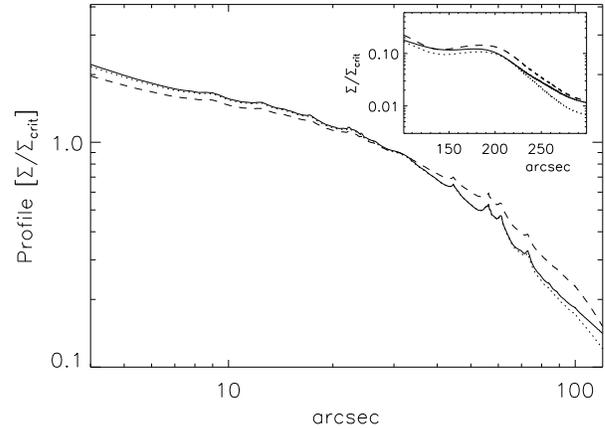}
   \caption{Convergence profiles for the 3 cases shown in figure \ref{fig_SLWL_CritCurves_5000_8000_15000}. 
            The solid line corresponds to the 5000 iteration case, the dotted line to the 8000 iteration case, 
            and dashed line to the 15000 iteration case. Again, the smaller sub-plot shows the tails of these 
            distributions across the transition phase and up the the $5\arcmin$ maximum radius.}
    \label{fig_Iter_M_C_profiles} 
\end{figure} 

\begin{figure}  
   \includegraphics[width=8cm]{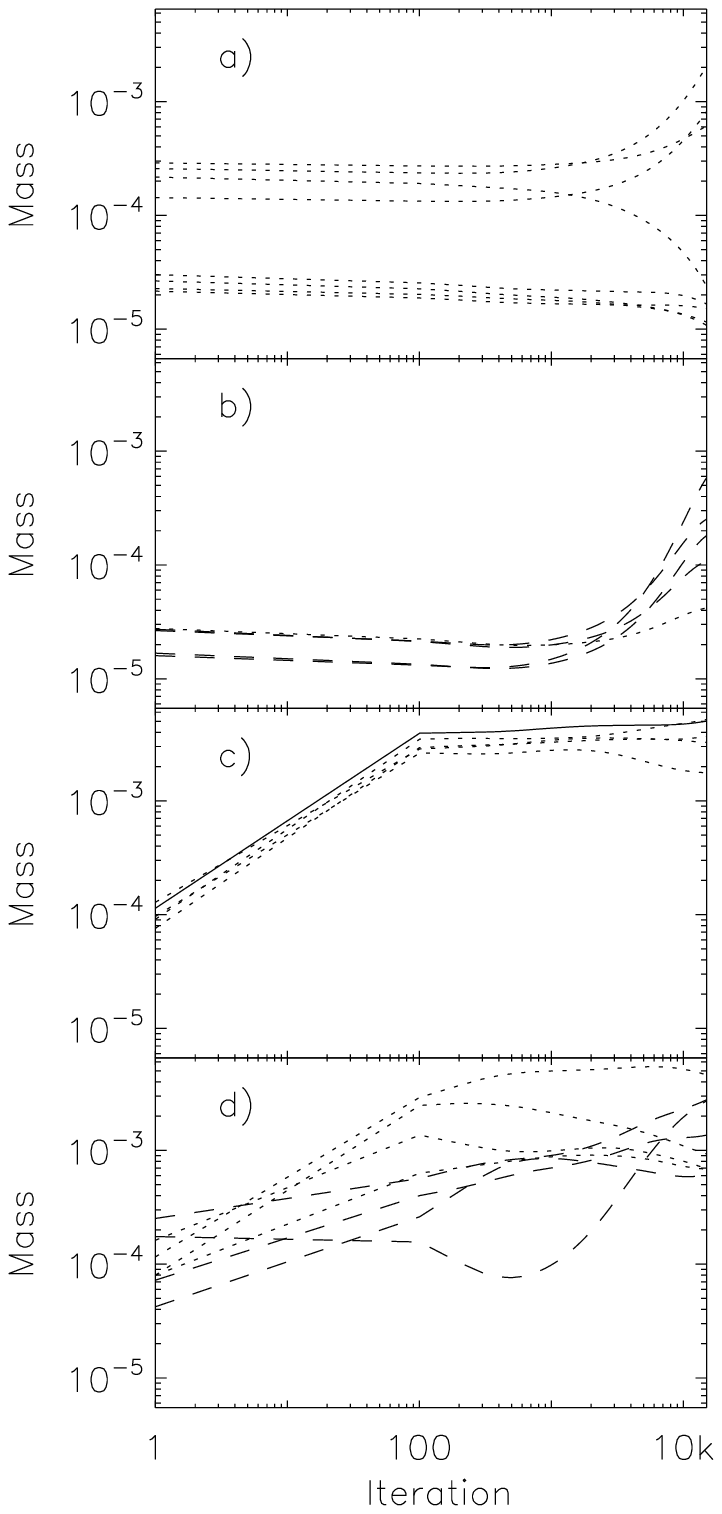}
   \caption{Evolution of the mass value (in units of $10^{15} M_{\odot}/h$ per cell) in specific cells with the 
            iteration number. 
            From top to bottom, panel a) shows the mass in the outer edge of the buffer zone. 
            The curves correspond to 4 contiguous 64x64 pixel cells 
            located at the bottom left corner of the 10x10 arcmin$^2$ field of view (bottom set of 4 curves) and 4 
            contiguous  64x64 pixel cells 
            in the opposite corner at the top-right corner of the field of view (top set of 4 curves). 
            Panel b) shows as a dotted line the case of a 64x64 pixel cell situated in the interior edge of the 
            buffer zone. 
            In particular this cell 
            is at the diagonal and at the transition phase between resolutions. 
            The dashed lines correspond to the 4 closest 24x24 pixel cells to the 64x64 pixel cell.  
            Panel c) shows with a solid line the cell situated in the centre of the field of view. The four dotted 
            lines are the 
            4 cells immediately to the right-left and up-down from the central cell. Panel d) shows the case of 4 
            cells situated at $0.5\arcmin$ from the centre 
            (dotted lines) and at $1\arcmin$ from the centre (dashed lines).}
    \label{fig_Iter_Cells}  
\end{figure}  

\section{Variability of the possible solutions}
%%%%%%%%%%%%%%%%%%%%%%%%%%%%%%%%%%%%%%%%%%%%%%%%%
We refer to the best fit model presented above as our {\it reference
solution} since in the context of this form of
modeling there is no single unique solution, given that the number
of lensed images and the grid resolution are finite. So more important
than finding a statistically {\it best} solution is to understand
the range of possible solutions that are consistent
with the data. Hence, as discussed in previous papers,
\citep{Diego05a,Diego05b,Diego2007,Ponente2011,Sendra2014}, we 
intentionally seek an approximate solution to the system of
equations. A major mistake in this form of modeling is to adopt a
grid of higher resolution than justified by the number of
lensed images, as then in the limit we may obtain what appear to be a
near perfect solution that matches identically the locations of all
lensed images but at the expense of a mass distribution that is much
more highly structured on small scales that is physically reasonable,
including negative surface densities. Such forced solutions tend to
predict sources (in the source plane) that are unreasonably small and
concentrated together in the centre of the field of view corresponding
to huge lens magnifications.  This over-fitting regime can be avoided by appreciating that
uncertainties in the data and the approximations made by our
non-parametric reconstruction means that a minimal, inevitable level
error must be allowed in the reconstruction. This includes our
assumptions and approximations introduced from our hypothesis 
that the member galaxy deflections are strictly proportional to the light, and
that the soft component can be exactly modeled by a superposition of
Gaussians of a given pixel scale, or that the sources in the source plane
are delta functions, or that there are no significant projection
of matter along the line of sight etc.

To allow for some error we may terminate the minimisation at a given point 
as described below and beyond which further iteration may result in unreasonably 
structured mass distributions. 
Since we are minimizing a N$_x$-dimensional quadratic function (where the number of dimensions
is the number of variables in the vector $X$), after a fixed number of
iterations (the number of iterations can be defined a priori) 
the algorithm stops in one of the infinite points contained
in the N$_x$-dimensional circumference at a height $\epsilon$ from the
minimum of the quadratic function.  The value of $\epsilon$ can be
estimated (and hence the maximum number of iterations) if we set a prior on
the mean size of the galaxies in the source plane and we combine this
information with the error in the shear measurements (see
\cite{Diego05a}).  We also generate simulated data-sets that mimic 
the data so that we can determine an optimal range for the number of iterations
that best reconstructs the input model, as described in
\cite{Sendra2014}. The starting point for the iteration process is
also important as different initial conditions may imply
the most reasonable point to stop iterating. 

As our input catalog  we consider first a reliable set of 26 robust strongly lensed
systems in which we have great faith. Some of these systems have
only photometric redshifts which can be imprecise. Changing the SL (or
WL) data set has an impact on the reconstructed solution as the
constraints in the system change accordingly. In this section we
explore these sources of variability. Other sources of variability
still exist including the number of grid cells to adopt or changing
the parameters that define the deflection field of the member galaxies.

\subsection{Dependency with iteration number}
%%%%%%%%%%%%%%%%%%%%%%%%%%%%%%%%%%%%%%%%%%%%%%%%%%

\begin{figure*}  
%   \centerline{ \includegraphics[width=16cm]{figs/A1689_Mean_SNR_Sigma.ps} }                                  
%   \centerline{ \includegraphics[width=6cm,angle=-90]{figs/Mean_Sigma_SNR_6p66arcmin_SL_WL10arcmin_Color.ps} }
   \centerline{
   \includegraphics[width=18cm]{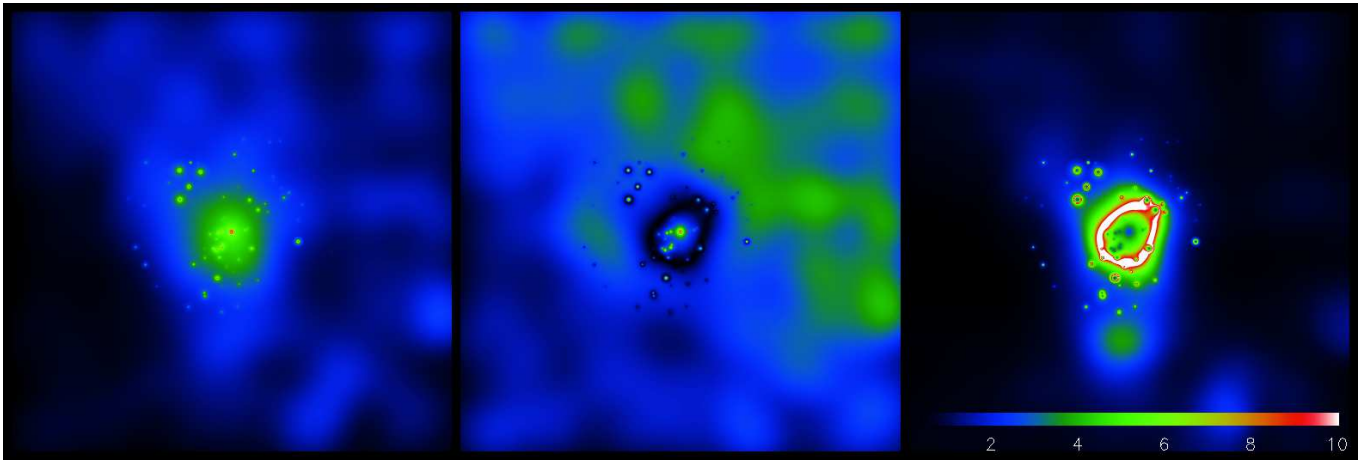}
   } \caption{ Reconstructed average mass from 40 independent
   solutions in the 10 arcminute field of view (left panel). 
   The middle panel shows the dispersion of
   these solutions.  The right panel shows the SNR map defined as the
   average map (left panel) divided by the dispersion map (middle
   panel). The SNR vary between SNR=0.7 at its minimum and SNR=290 at
   its maximum (but saturated in this plot above SNR=100). All maps
   are shown as the square root (in order to increase contrast) and
   the color scale (shown in the right panel) is the same in all
   panels.}  \label{fig_mean_SNR_sigma}
\end{figure*}  

The maximum number of iterations chosen determines
some of the properties of the solution.  In \cite{Sendra2014} we
discussed how for a simulated data set of SL measurements that
was designed to resemble the real data of A1689, the optimal number of iterations
was of the order of several thousands.  Also, an interesting
conclusion form that work is that by incorporating the deflection
field of the galaxies, one gets the added bonus of increasing the
stability of the solution which tends to saturate to a fixed minimum level of precision 
as the number of iterations becomes very large. The over-fitting problem was reduced as well
but nevertheless,  over-fitting can occur if the number of iterations is too
large so the algorithm always needs to be stopped after a given number
of iterations. For this purpose, we have found that using the location of the radial
critical curves is a sensitive choice for  identifying a sensible range for the number of
iterations, as the radius of this critical curve is well defined in the data as it
can be seen to be fairly circular in shape from the distribution of very 
radially extended images. This is not the case for the tangential critical curve.

In this paper, this part of the analysis is done a
posteriori (although it could be in principle incorporated into the
system of linear equations, however, this is beyond the scope of this
work and will be studied in more detail in a future paper).  
In figure \ref{fig_SLWL_CritCurves_5000_8000_15000} we show
three solutions corresponding to three different numbers of iterations
(where the number of parameters fitted and the initial condition in  the minimization 
are identical between these three solutions). From left to right we show the solutions obtained after
5000, 8000 and 15000 iterations. By looking at the tangential curves,
the three cases look different but they still accommodate well the
large tangential arcs in between the tangential critical curve. The
radial curve, on the other hand, when compared with the position of
key well identified radial arcs, is clearly too large in the case of
5000 iterations and too small in the case of 15000 iterations. In
contrast, the case of 8000 iterations, the radial critical curve
(computed for z=2) overlaps almost perfectly with the mean position of 
different radial arcs present in the cluster at redshifts $z \approx 2$. This simple
comparison seems to indicate that the best solutions are obtained with
our code after $\sim 8000$ iterations. Using the radial critical curve as a way of 
determining the optimal range of iterations can be also seen as a regularization of 
our problem. This is an interesting alternative since it is solely based on actual data.

An idea of the dependency of the solution with the number of iterations can be obtained 
also from figure  \ref{fig_Iter_M_C}. In the top part of this plot we represent the total mass 
contained in the grid (solid line)  versus the iteration number. The solid and dashed lines 
represent the correction factors ${\rm C}_1$ and ${\rm C}_2$,  that are applied to the fiducial 
deflection fields from the galaxies. As the iteration number grows, there starts to be a trade 
between the mass contained in the grid, and the mass in the galaxies but the total mass (especially 
in the  central region) stays more or less constant beyond a few thousand iterations as can be 
seen better from the profiles  (figure \ref{fig_Iter_M_C_profiles}). A similar trend was observed 
when applying the method to simulated data \citep{Sendra2014}. As the iteration number grows, the 
solution increases its complexity in order to concentrate the arcs into smaller sources. As the fiducial 
field has only two degrees of freedom ($C_1$ and $C_2$), new features appear only in the grid part of 
the solution at the expense of reducing the mass in the member galaxies to keep the total mass more 
or less constant (within the Einstein radius). Of course, and as mentioned earlier, to avoid over-fitting 
the minimization process has to be stopped at some point (stopping the minimization after a number of 
iterations could be seen also as a regularization process). In the range of iterations (6000-10000) 
where reliable solutions exist, the changes in the model are small in the central 4 arcminute region. 
It is also important to note that the increased raise in mass in the grid part of the solution after 2000 iterations 
(see figure \ref{fig_Iter_M_C}) is driven mostly by the cells in the outermost region 
(beyond the Einstein radius, as shown more clearly below in figure \ref{fig_Iter_Cells}). 
It can also be seen (see the inset  in the upper right corner) that as the iteration number grows, the grid starts to 
accumulate mass in the transition phase between the $6.66\arcmin$ region and the buffer zone. 
The bottom panel of figure \ref{fig_Iter_M_C} shows the quantity that is being minimized (properly 
re-scaled by some constant for clarity) as a function of the iteration number.  

\begin{figure}  
%\centerline{ \includegraphics[width=8cm]{figs/Mass_profile_A1689_IT5000_34src_2ly_SLWL_Nmin20.ps} }                     
\centerline{ \includegraphics[width=9cm]{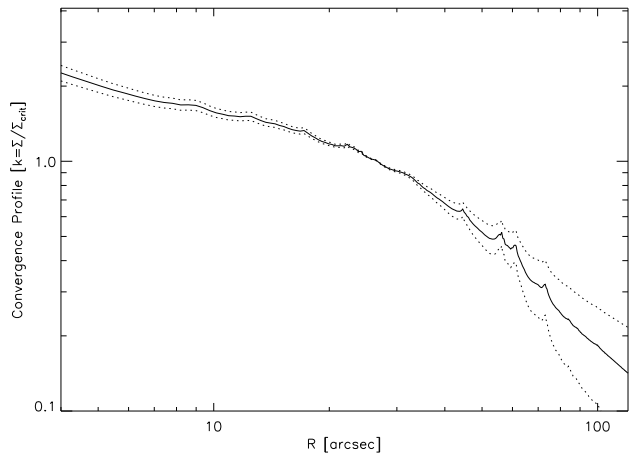} }                     
   \caption{  
            Average profile of the convergence from 40 independent solutions (solid line). The dotted 
            lines represent the dispersion of these solutions.  
           }  
   \label{fig_profile_sigma}  
\end{figure}

An alternative (and illustrative) way of looking at the evolution of the solution with the iteration number is 
shown in figure \ref{fig_Iter_Cells} where we show how the mass in individual cells evolve with the iteration number. We choose cells that are representative of a larger region. The top panel shows the typical behaviour of cells in the outer border of the buffer zone (they correspond to two diagonal cells at more than $5\arcmin$ distance 
from the centre of the field of view). During the first few hundred iterations, the mass in these 
cells change very  little but they become more {\it active} at later times when the solution 
at the central region has been constrained 
and remains more stable. As the iteration number grows, some cells in this region 
gain mass and some others lose it. Panel b) shows what happens at the transition phase between the two resolutions. Again, 
this region plays a less important role at the beginning of the minimization but they become more {\it active} before 
the cells in the outer region. Also, it is interesting to see how both the large cell (dotted line) as well as the 
4 neighbouring smaller cells (dashed lines) evolve simultaneously and also follow the same trend (increasing their 
mass). As can be seen from 
this figure and figure \ref{fig_Iter_M_C_profiles}, the transition phase tends to accumulate mass as the iteration 
number grows. Panel c) shows what happens to the most central cells. The cell occupying the centre of the field of 
view corresponds to the solid line and the dotted lines correspond to the 4 cells which are closest to the central one. 
In this case the value on these cells converge much faster indicating that the algorithm naturally tends to put mass 
in the centre first and then moves outwards. This behaviour is not imposed but just happens naturally. 
The cells in the centre don't reach a stable point but neither does the $C_1, C_2$ parameters shown in 
figure \ref{fig_Iter_M_C}. Meanwhile, the profile seems to be less sensitive to the iteration number (see figure 
\ref{fig_Iter_M_C_profiles}) indicating that there must be some trade-off mass between the grid and galaxies in order 
to compensate each other and keep the profile stable. Finally, panel d) shows an intermediate region between the 
centre of the field of view and the buffer zone. In this case, the evolution of the cells is more complex with no 
clear tendency. As in the case of the very central cells, the ones in this region evolve faster than the more distant 
ones (although not as fast as the most central cells). However, there seems to be more spatial variability in 
this region than in the centre and also less tendency to converge, especially those cells located beyond the Einstein 
radius (dashed lines) which at later times become more {\it active} under the influence of the WL data.

\subsection{Dependency with the initial guess}\label{subsect_20min}
%%%%%%%%%%%%%%%%%%%%%%%%%%%%%%%%%%%%%%%%%%%%%%%
\begin{figure}  
   \includegraphics[width=8cm]{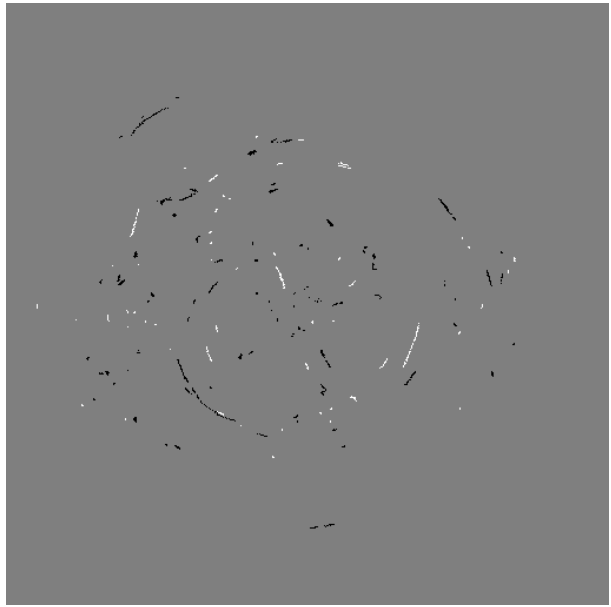}
   \caption{In black the original robust data set of 26 sources. 
            In white we show the additional arcs that together with the previous 26 sources 
            data set conform the extended 48 source data set of table 2 (sources 25 and 32 are not used). 
            The field of view is $3.33\arcmin$.}
    \label{fig_Arcs_26vs49src}  
\end{figure}  

\begin{figure}  
   \includegraphics[width=9cm]{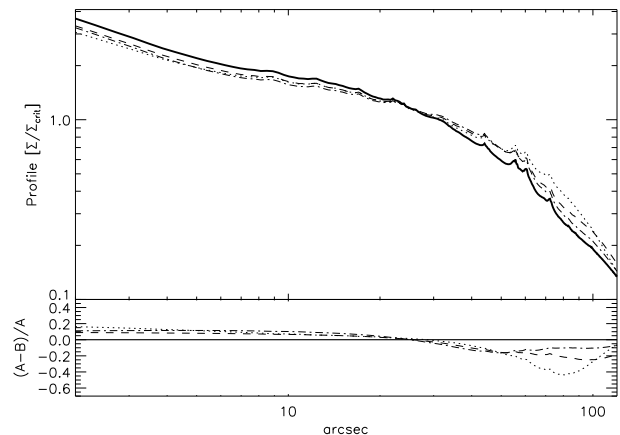}
   \caption{Comparison of the profiles obtained with different subsets of sources. The thick solid line 
            corresponds to our reference solution obtained with the robust subset of 26 sources. 
            The dotted line corresponds to the solution 
            obtained with the full sample of 48 sources in table 2 (we exclude sources 25 and 32), or case i) (see text). 
            The dashed line corresponds to the subset of 18 sources with spectroscopic redshifts or case ii) (see text) 
            and the dot dashed line corresponds to the same initial set of 26 sources but the sources 
            with photo-z taking substantially different redshifts, or case iii) (see text).}
    \label{fig_Profiles_26_49_18_26b}  
\end{figure}  
When minimizing a multi-dimensional quadratic function one can find infinite solutions (for a given error)
by varying the starting point of the minimization, all of them equally good in the sense of fitting the data set. 
The regions in the lens plane that are more sensitive to the data will converge quickly towards stationary 
points while regions with weaker sensitivity to the data might vary more 
from minimization to minimization or even not vary significantly and retain values close to their initial values 
(memory effect).  

In figure \ref{fig_profile_sigma} we show the average (solid) and 1-sigma region (dotted) of 40 reconstructions 
where in each one we change the initial condition by setting it to a vector of random numbers obtained from a 
Gaussian distribution. The dispersion of this Gaussian distribution is a random variable itself and is such that the 
total mass in the initial condition takes values in the range $[\sim 0.2,\sim 2]\times 10^{15} M_{\odot}$/h. 
Table 1 includes the values of the mean and errors as a function of distance. 
The solution retains some memory of the initial condition specially in the outer region 
where the constraints in the solution are the weakest. The most remarkable aspect of this plot is that all solutions 
seem to converge to the same profile in the range $20\arcsec-30\arcsec$ defining a {\it stability} 
region for the solution. 

    \begin{table}
      \centering
    \begin{minipage}{55mm}                                               
    \caption{Mean profile and dispersion of the 40 solutions shown in figure \ref{fig_profile_sigma}. 
             Both the mean profile and dispersion are given in units of $\Sigma/\Sigma_{crit}$ with 
             $\Sigma_{crit}$ computed at $z=1.07$. The last column shows the integrated (cylindrical) 
             mass in units of $10^{14} M_{\odot}/h$.}
 \label{tab2}
 \begin{tabular}{cccc}   
  \hline
   arcsec  &    Mean   &   Disp. &   M$(<r)$  \\
  \hline
    2.343  &    3.117  &  0.282  &  0.015 \\
    4.296  &    2.174  &  0.149  &  0.036 \\
    7.031  &    1.744  &  0.089  &  0.072 \\
    10.15  &    1.562  &  0.064  &  0.128 \\
    12.50  &    1.514  &  0.058  &  0.179 \\
    14.45  &    1.411  &  0.046  &  0.228 \\
    16.40  &    1.327  &  0.036  &  0.281 \\
    18.35  &    1.242  &  0.026  &  0.339 \\
    20.70  &    1.153  &  0.016  &  0.412 \\
    23.04  &    1.133  &  0.015  &  0.492 \\
    25.78  &    1.020  &  0.003  &  0.592 \\
    28.12  &   0.9477  &  0.006  &  0.680 \\
    30.46  &   0.9079  &  0.010  &  0.772 \\
    32.81  &   0.8616  &  0.015  &  0.869 \\
    35.15  &   0.7836  &  0.024  &  0.968 \\
    37.89  &   0.7198  &  0.032  &  1.083 \\
    40.62  &   0.6644  &  0.039  &  1.201 \\
    43.35  &   0.6308  &  0.044  &  1.320 \\
    46.09  &   0.5962  &  0.049  &  1.444 \\
    48.82  &   0.5367  &  0.057  &  1.565 \\
    51.56  &   0.4985  &  0.062  &  1.683 \\
    54.68  &   0.4976  &  0.063  &  1.819 \\
    57.81  &   0.4616  &  0.067  &  1.961 \\
    60.93  &   0.4619  &  0.067  &  2.097 \\
    64.06  &   0.3747  &  0.077  &  2.226 \\
    67.18  &   0.3324  &  0.081  &  2.341 \\
    70.70  &   0.3148  &  0.082  &  2.466 \\
    74.21  &   0.2976  &  0.081  &  2.589 \\
    77.73  &   0.2592  &  0.083  &  2.702 \\
    81.25  &   0.2402  &  0.083  &  2.808 \\
    84.76  &   0.2292  &  0.082  &  2.910 \\
    88.28  &   0.2145  &  0.080  &  3.011 \\
    92.18  &   0.2007  &  0.079  &  3.119 \\
    96.09  &   0.1900  &  0.078  &  3.225 \\
    100.0  &   0.1837  &  0.077  &  3.350 \\
    103.9  &   0.1725  &  0.076  &  3.432 \\
    107.8  &   0.1638  &  0.076  &  3.530 \\
    112.1  &   0.1554  &  0.075  &  3.634 \\
    116.4  &   0.1474  &  0.074  &  3.735 \\
    120.7  &   0.1409  &  0.074  &  3.829 \\
    125.3  &   0.1341  &  0.073  &  3.929 \\
    130.0  &   0.1285  &  0.072  &  4.040 \\
    135.1  &   0.1238  &  0.071  &  4.125 \\
    140.6  &   0.1204  &  0.070  &  4.234 \\   
  \hline
 \end{tabular}
 \end{minipage}
\end{table}

\begin{figure*}  
   \includegraphics[width=5.8cm]{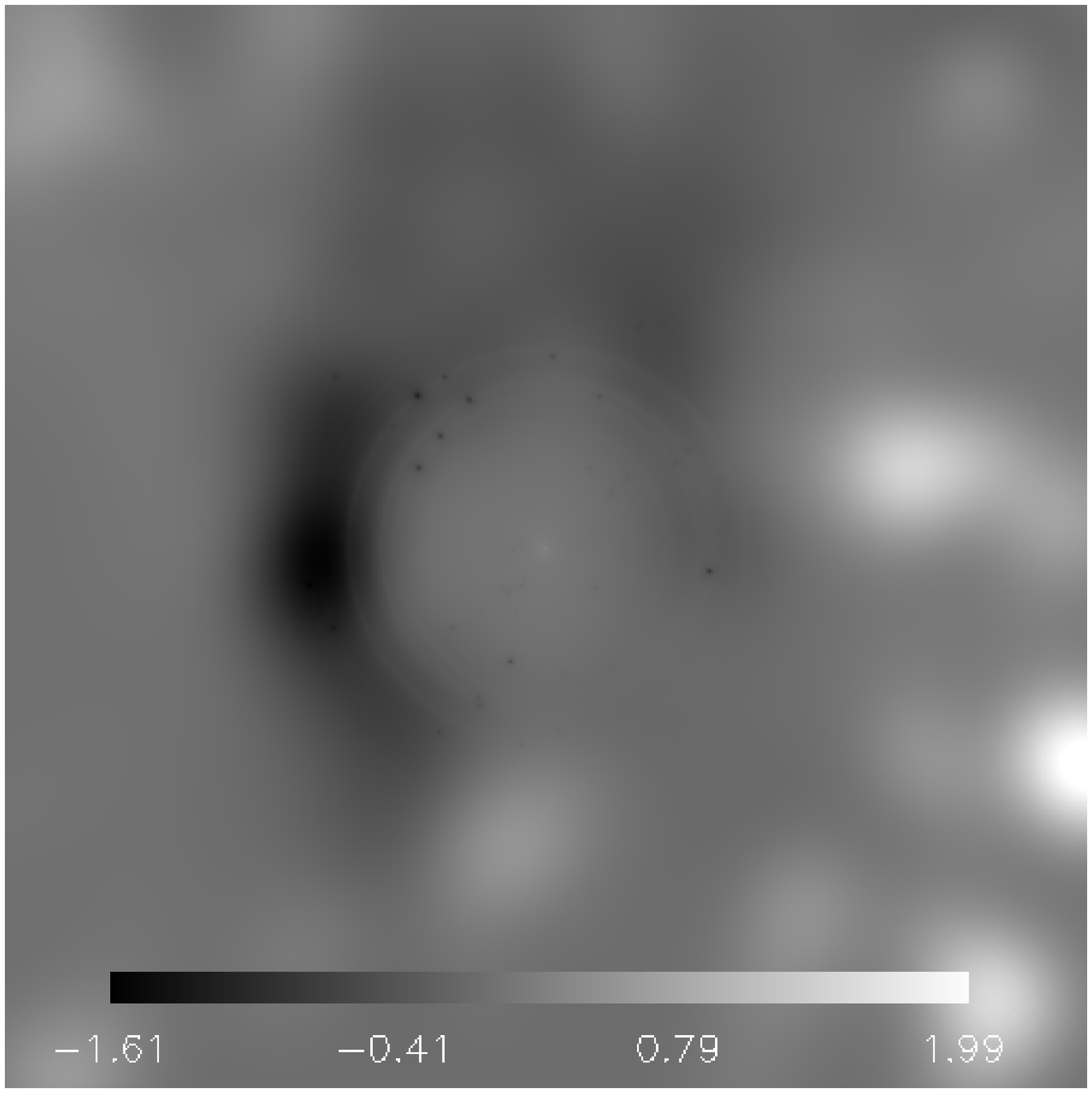}   
   \includegraphics[width=5.8cm]{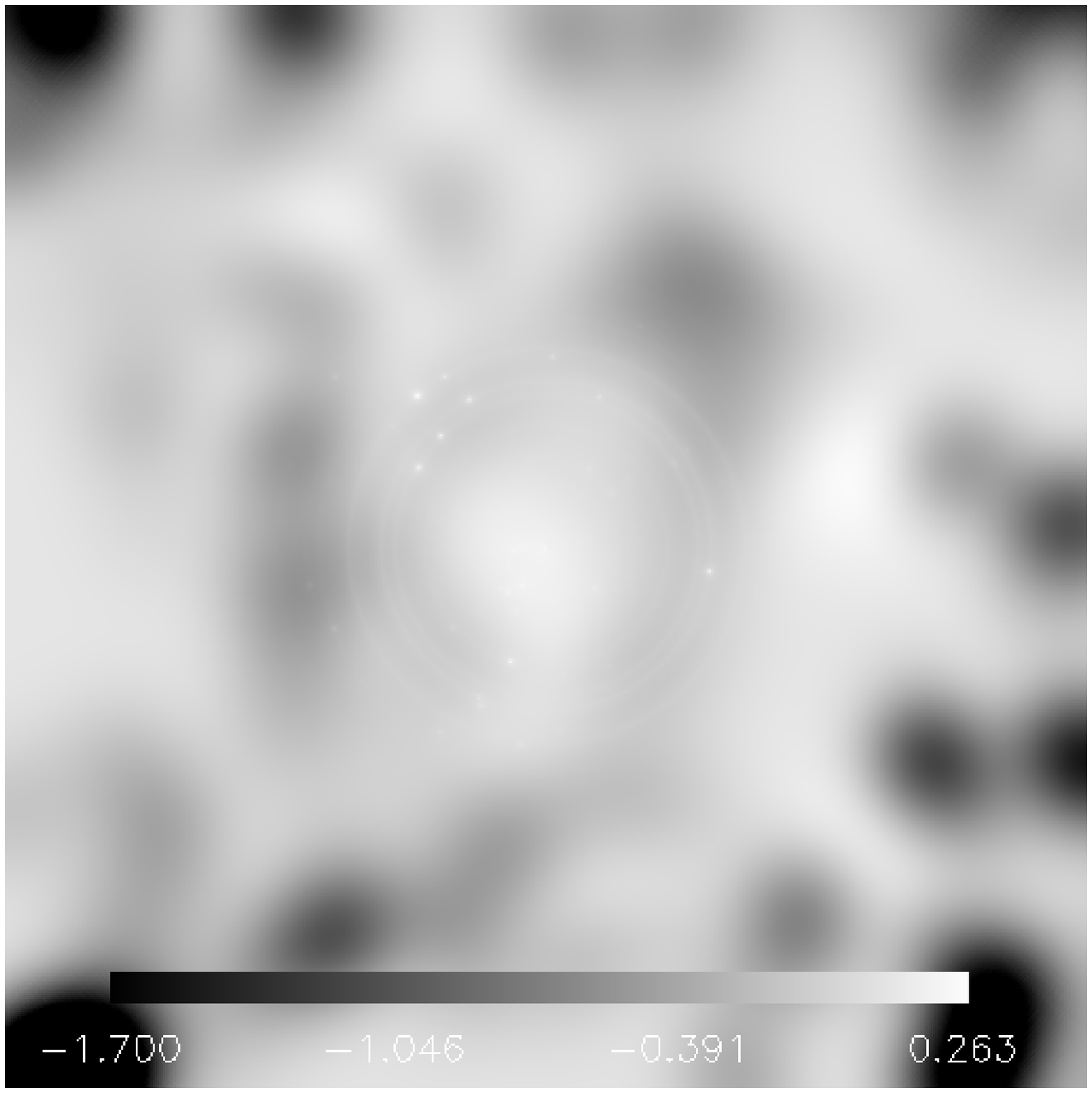}   
   \includegraphics[width=5.8cm]{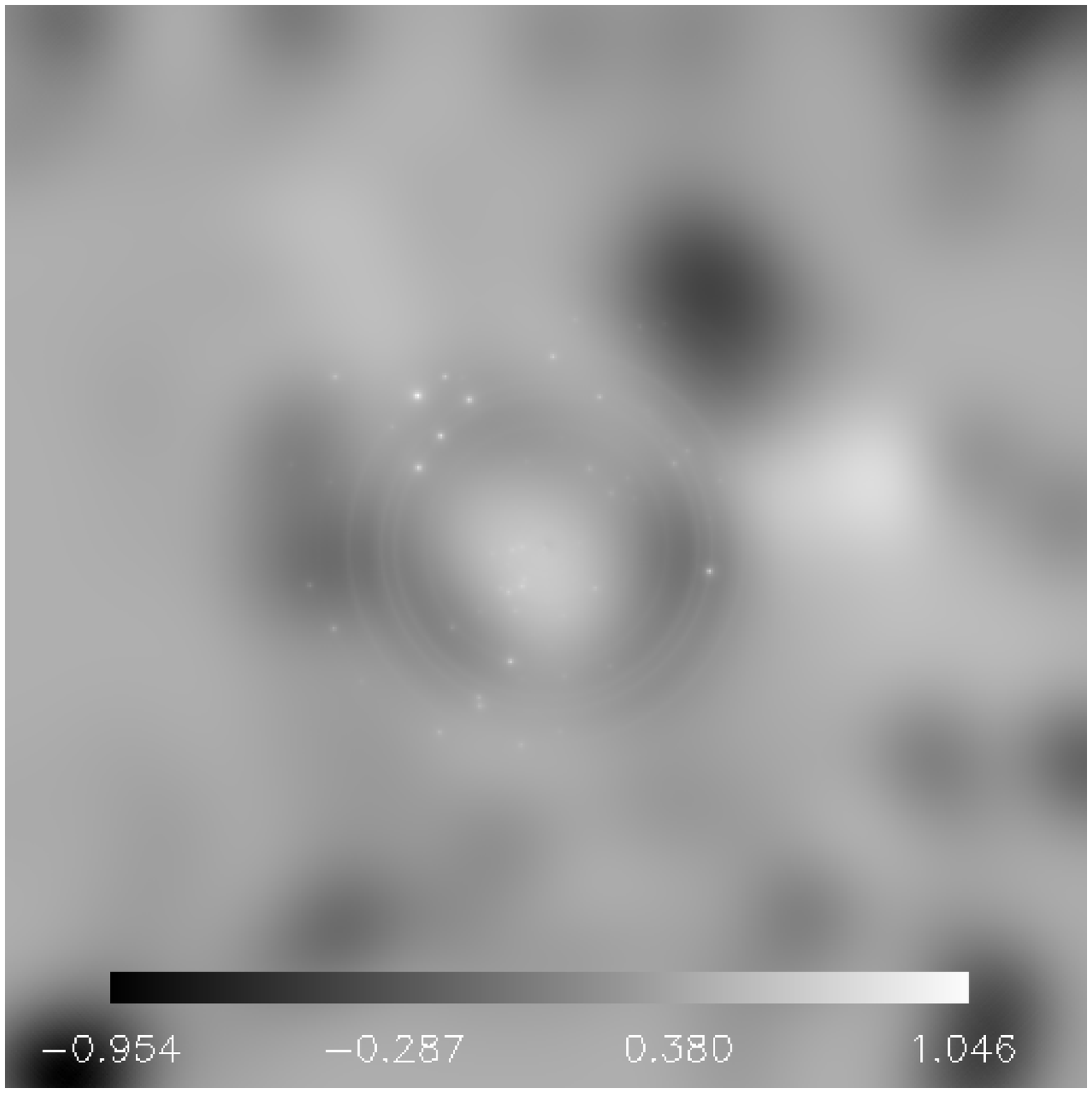}   
   \caption{The figures show the difference between our reference solution and solutions obtained with three 
           alternative subsets of sources. In all cases, the difference has been divided by the profile of the 
	   reference solution in order to visually maximize the differences. Hence a value of 2 in this graph 
           means that in the 
           difference of the two solutions (reference-alternative) there is 2 times the average mass of the profile 
           for that radial bin etc. Some values (near the edges) have been saturated also for contrast purposes 
           (the middle panel had a large negative deviation of -5 in the bottom left corner that was saturated to -1.7). 
           The field of view corresponds to $6.66\arcmin$ across.
           Left panel: case for which all 48 sources in table 2 are used (we exclude 25 and 32). Middle panel: case for which only the 
           spectroscopic sub-sample of 18 sources is being used. Right panel: case for which the original 26 
           sub sample of sources is used but the sources with photometric redshift have different redshifts  
           }  
   \label{fig_MassDifference2}  
\end{figure*}

This convergence is made more evident when looking at the 2D version of the above result. 
In figure \ref{fig_mean_SNR_sigma} we show the average of the mass solutions (left panel) 
from which the above profile is derived. The middle panel shows the 
dispersion of the mass map and the right panel shows the signal-to-noise ratio (or SNR) defined as the ratio of the 
left and middle panels. There is a remarkably well defined circular region of high SNR around $20\arcsec-30\arcsec$  
from the centre. In this region, the solutions seem to be insensitive to the initial condition and they all render almost 
identical results. 
This region corresponds to a stability region of the solution where the profile is constrained very well. 
A second interesting aspect can be seen also in the SNR map. The high SNR of the dark matter blob, 
just south of the stability region, indicates that this might be a real substructure. This feature is not previously
claimed and indeed parametric methods would not incorporate such dark sub-structure in principle 
and hence this is the first time it is revealed.  
Other interesting features emerge from the SNR map as potential real substructures of the cluster although with a 
lower SNR. The dispersion map shows how there is a region (upper-right quadrant) where the solutions fluctuate the most 
indicating that the mass map is less reliable in this region. 
 
Finally, as expected (and already shown in the profile plot), the dispersion of the solutions in the central 
region (that is the cD galaxy) is larger, indicating therefore that our solution is not very sensitive 
to the very central region. This is due mostly to the fact that the larger arcs have a bigger weight in our solution. 
In a future paper we aim to study in more detail the most central region taking advantage of the stability region 
identified in this paper, and the possible implications for the mass profile of the cD galaxy.

\subsection{Dependency with the number of systems and redshift}
%%%%%%%%%%%%%%%%%%%%%%%%%%%%%%%%%%%%%%%%%%%%%%%%%%%%%%%%%%%%%%%%

Identifying pairs of images in the strong lensing regime is not always free of subjectivity, resting on experience.
A1689 is probably the most scrutinised  lens, with hundreds arclets seen in the image, each 
with its own distortion that makes it difficult to find morphologically similar galaxies. 
The colors may also vary across the object due to differential magnification, and also because of overlap 
with other unrelated images or due to diffuse light in the cluster affecting the colours of relatively faint 
background galaxies. Spectroscopic redshifts of the arclets are often the best way to discriminate among different  
possibilities but even this does not always settle the differences and furthermore neighbouring galaxies may 
become confused in the process if they have the same redshift to within the resolution limits. A good example of this 
is the difficulty to distinguish between systems 10 and 12 (both having basically the same spectroscopic redshift). 
Different authors have assigned the arclets to different systems (see table 2). In this section we explore the 
impact on the solution when we consider different subsets of arclets in our SL part of the data set. 

In  table 2 we compile all the systems that where found in the 
literature and we add several new systems (candidates) which are identified with our model.
For the new systems we simply assume that they are at redshift $z=2$ when assessing their deflection angles. 
This is a useful approximation given the weak dependency of the deflection angles  over the range $z = [1,3]$ 
(as shown by figure \ref{fig_fk}) and should be sufficient for our purposes. 
For our test we  compare our reference solution described in section \ref{sect_results}, 
which depended on only the 26 established systems, with new model solutions obtained for sets of images 
composed in the following 3 ways:

(i) We consider an extended sample of systems listed in table 2 but exclude sources 25 and 
    32 for which multiple options exist for the same system. That is, we consider 48 sources out of the total 50 
    systems listed  in table 2.\\
(ii) As a second sub-sample we consider the subset of 18 sources from the original 26 sources of our reference 
     solution for which spectroscopic redshifts are available, in table 2. \\
(iii) The third sub-sample is the same as the original sample of 26 but for the sources with photometric redshifts 
      we allow the redshifts to vary by a generous $2\sigma$ error, adopting the $\sigma$ values in \cite{Coe2010}.\\

The extended data set of 48 systems is displayed in figure \ref{fig_Arcs_26vs49src} where the original 
26 systems are shown in black and the additional 22 systems in white. For each case, we reproduce 
the minimization process of section \ref{sect_results} that is, we adopt the same initial 
condition, number of iterations and make use of both  SL and WL data, in order to better examine 
model differences resulting from changes to the input the data set. Figure \ref{fig_Profiles_26_49_18_26b} shows 
this comparison in terms of the resulting mass profiles for the above cases.

\begin{figure}  
   \includegraphics[width=8cm]{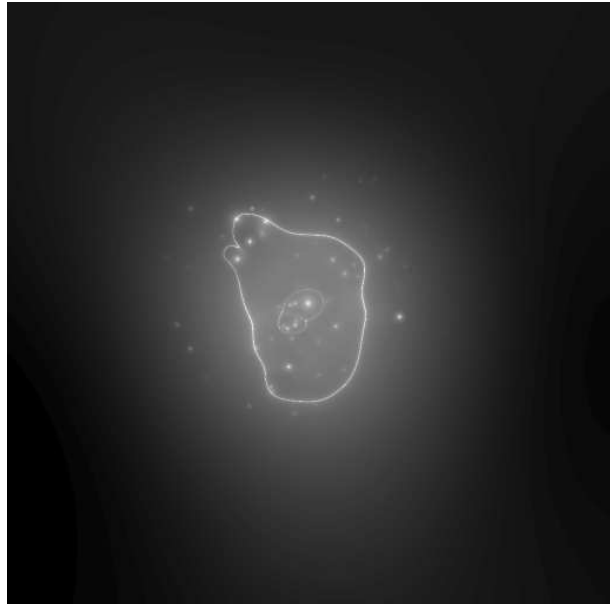}
   \caption{Mass and critical curve for a low resolution grid reconstruction (SL+WL). 
            The cell sizes are 2.7 times larger in this case than 
            in the reference solution.}
   \label{fig_CritCurve_RG}  
\end{figure}

\begin{figure}  
   \includegraphics[width=8cm]{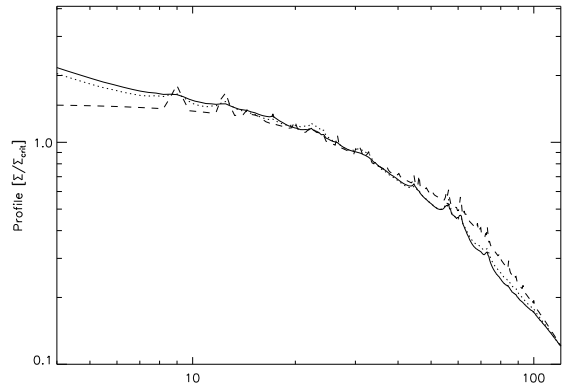}
   \caption{Change in reconstructed profile under three assumptions for the galaxies in the cluster. The solid 
            line corresponds to the reference solution discussed above, the dotted line is for a different model where  
            both the scale radius and total mass of the assumed NFW profile for the individual galaxies is changed by 
            a factor $\sim 2$. The dashed line corresponds to another different realization of the masses in the 
            galaxies (different also by a factor $\sim 2$ with respect to the reference model) but their profiles 
            are taken to the extreme case of delta functions. For the dashed line, there is a peak 
            at the centre not shown in this plot that corresponds to the central galaxy.}
   \label{fig_profiles_Rs}  
\end{figure} 
  
Although there are some differences between the different solutions, the agreement is still remarkably good indicating 
that all data sets have enough common systems to produce similar results and/or that the solution is not very sensitive 
to modest changes in the redshift of some systems. This agreement indicates most simply that  the systems in the 
extended sample naturally give good fits when the reference solution is used. 
The last point is highlighted better in the last column of table 2 where we show 
the $\Delta \beta$ for each arclet (and based on the reference solution). 
The  $\Delta \beta$ for a specific arclet $i$ is defined as,
\begin{equation}
\Delta \beta_i(arcsec) = |\beta_i - <\beta>|
\label{eq_chi2} 
\end{equation}
where $\beta_i$ is the predicted position of the arclet in the source plane when the reference solution is used and 
$<\beta>$ is the average of all the $\beta_i$ for that system.
As shown by the values of $\Delta \beta$ (expressed in arcseconds), 
most arclets lie at reasonable distances (few arcseconds) 
from their common centre in the source plane. Also, some systems appear to be problematic since they have large 
 $\Delta \beta$ values ($10\arcsec$ or more, like in systems 25, 41, 44). 
These systems are either incorrect  or our reference solution has substantial errors around the 
position of these systems. In other cases, like system 7, the arclet 7.3 ($\Delta \beta = 9.3\arcsec$) is very close to 
the central cD galaxy and surrounded by multiple small sources (7.3, together with 8.5, and 19.5 where also rejected 
by \citep{Coe2010}). It is possible that either the reference solution is not 
accurate at the centre (see discussion in section \ref{subsect_20min}) or that 7.3 does not correspond with the source 
listed in table 2 but another one in the vicinity. If that's the case, this explains why 7.1 and 7.2 
(which are clearly the same source), have also relatively high values of $\Delta \beta$ 
since a bad association for 7.3 would bias $<\beta>$ and enhance the  $\Delta \beta$ for all the system. 

Another interesting point from figure \ref{fig_Profiles_26_49_18_26b}, is that the same stability region discussed 
in section \ref{subsect_20min} seems to be present when we change the SL data set, in the range $20\arcsec-30\arcsec$,
where the solutions seem to be insensitive to the particular choice of data sets (among the 4 used in this comparison). 
This reinforces the idea that the solution is very well constrained in this regime ($20\arcsec-30\arcsec$) and nearly 
insensitive to the intrinsic variation of the solutions. 

Going beyond the differences in profile, we look at  the mass maps for the above cases, comparing the reference 
solution and the 3 solutions described above. The result is shown in figure \ref{fig_MassDifference2}. 
In this case, the mass difference has been divided by the profile of the reference solution to increase contrast. 
Also, the middle panel has been saturated below values of -1.7 (the most negative value was -5) also for contrast 
purposes.  The difference maps show where the surface mass density modifies itself in order to accommodate the 
possible changes in the data set and by extension, it marks the regions where extra caution needs to be taken 
into account when interpreting our main results, should the assumed original data set of 26 sources be compromised 
by systematics.

\subsection{Dependency on the grid configuration}
%%%%%%%%%%%%%%%%%%%%%%%%%%%%%%%%%%%%%%%%%%%%%%%%%%
The choice of grid resolution is an important decision that affects the performance of the reconstruction 
and must be made with some care.
Ideally we would set a large number of cells that allows for a more detailed reconstruction. In practice however, 
a very large number of cells results only in a more noisy reconstruction that needs to be smoothed (regularized). 
The smoothed image captures the main eigen-modes of the solution that would be reconstructed also with a smaller 
number of cells. Also, a large number of cells implies a larger system of equations to resolve and a 
slower convergence. This limits the ability to explore the space of possible solutions which is important. 
On the other hand, a very small number of cells results in a more compact system of linear equations that 
can be resolved fast but at the expense of not capturing some of the potentially smaller scale details of the 
mass distribution and consequently forcing the entire mass distribution to adopt, often erroneous, distributions 
in order to fit the observations. As shown by \cite{Ponente2011,Sendra2014}, a low resolution grid can still 
produce reliable solutions but only after allowing for a larger error in the reconstruction. 
Figure \ref{fig_CritCurve_RG} shows the reconstructed solution when a lower resolution grid is used. 
In this case, the cell sizes are 2.7 times larger than the cell sizes in the reference solution.  
The solution resembles a smoothed version of the reference solution 
except in the radial critical curve where the lack of resolution does not allow for a detailed reconstruction. 

A compromise alternative would be to use a multi-resolution grid where regions with a complex mass 
distribution are sampled with smaller cells and 
regions with a more smooth mass distribution (or less sensitive to the data) are sampled with larger cells. 
This possibility was explored in previous papers and produces satisfactory results although it is not entirely  free 
of problems. In particular, the boundary region between regions of differing resolution 
tends to produce biased results (when tested with simulated data). 
The size of the cell normally introduces a prior in the solution. The reconstruction tends 
to put more mass in the smallest cells. This problem can be mitigated by solving in an iterative way where the first 
iteration assumes a regular grid (no prior) and the consecutive iterations increase the resolution in regions where the 
previous iteration found more mass. 
We have tried a battery of configurations and found that some of them (in particular 
those with multi-resolution) produce significant artifacts in the solution that need to be avoided. 
Regular grids are always more stable and reliable so in our particular case and due to the large field of view 
involved (10x10 arcminutes$^2$) we choose a grid with good resolution in the central field of view (6x6 arcminutes$^2$) 
and a lower resolution grid in the outer region (buffer zone). 
As expected, the solution near the boundary between the two regions produce artifacts as discussed above. 
We have checked that variations in the grid sizes of $\sim 30\%$ with respect to our configuration, finding the 
results are nearly identical to the those presented in this paper. 
The cell sizes in the central part of the field of view are larger than the halo sizes 
of the galaxies (except the central galaxy) facilitating the orthogonality of the grid+galaxy base.

\subsection{Dependency with the parameters in the fiducial model}
%%%%%%%%%%%%%%%%%%%%%%%%%%%%%%%%%%%%%%%%%%%%%%%%%%%%%%%%%%%%%%%%%%%
In order to test the sensitivity of the solution with the assumptions made for the adopted profile for the member 
galaxy component we compare the reference solution (solid line in figure \ref{fig_profiles_Rs})
with two different assumptions regarding galaxy profiles. 
In the first case  (dotted line in figure \ref{fig_profiles_Rs}), we vary both the total mass of each galaxy 
and the scale radius by a factor $\sim 2$ (above and below the values in the reference solution). 
In the second case, (dotted line in figure \ref{fig_profiles_Rs}), we take the extreme (and unrealistic) case 
where all the masses in the galaxies are considered to be in just the centre of each galaxy 
(that is, galaxies are considered delta functions) and we also change their masses by a factor $\sim 2$ with 
respect to the reference solution case.  Remarkably, there is a region of stability (around $20\arcsec-40\arcsec$) 
where the profile seems to be unaffected by the particular choice of the fiducial deflection field, 
even in the case where we consider unphysical assumptions for the fiducial field (delta function case for 
the galaxy masses). The dotted line (different NFW profiles for each galaxy) is almost indistinguishable 
from the reference solution. The dashed line case (delta functions) shows a relative deficit in mass at 
very short radii, since all the mass of the central galaxy is concentrated at ($r=0$) which is not shown in the plot. 
At larger radii ($r=50\arcsec-100\arcsec$), the grid part of the solution shows and excess of mass (with respect 
to the reference solution). This might be due to a compensation effect of the unphysical nature of the 
delta function assumption although this is not observed at $r=20\arcsec-30\arcsec$. 
When finding solutions we are of course dealing with the deflection field and since this relates to 
the gradient of the potential we may not be so surprised about the independence of the resulting model
mass distribution with the choice of galaxy profile, since the galaxy potential is always extended even in the 
case of point masses. What is much more important here is that there is a contribution to the deflection field 
at the location of the member galaxies rather than the definition of the member galaxy mass profiles.

%%%%%%%%%%%%%%%%%%%%%%%%%%%%%%%%%%%%%%%%%%%%%%%%%%%%%%%%%%%%%%%%%
\section{Predicted new systems/arcs} \label{sect_newarcs} 
%%%%%%%%%%%%%%%%%%%%%%%%%%%%%%%%%%%%%%%%%%%%%%%%%%%%%%%%%%%%%%%%%
\begin{figure}  
   \includegraphics[width=8.2cm]{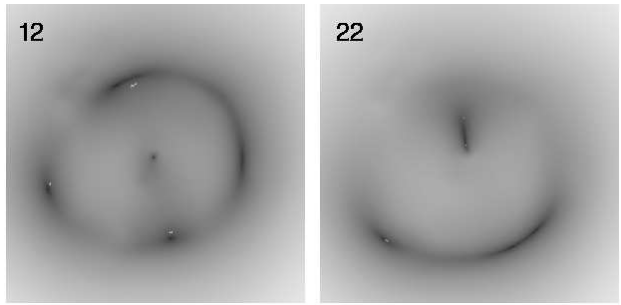}
   \caption{Predicted positions for systems 12 (left) and 22 (right). 
            Data (arcs) are shown in white and model prediction in dark grey. 
            For system 12, the solution correctly predicts an arclet (on the right side) 
            identified by Limousin et al. (2007). 
            For system 22, no counter-image is found at or near the position predicted by the 
            solution (bottom right).}
   \label{fig_predicted_arcs}  
\end{figure}

\begin{figure*}  
   \includegraphics[width=16cm]{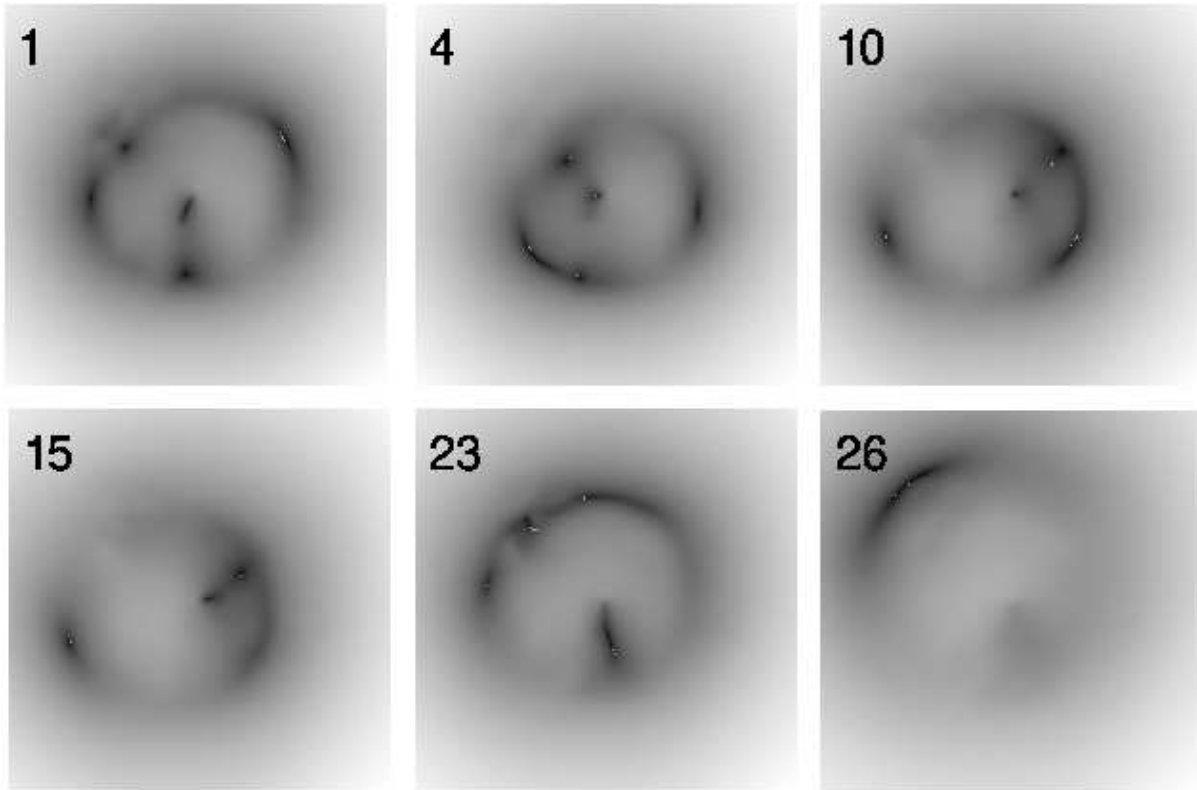}
   \caption{Some examples of systems where the model accurately predicts the arc positions. Most systems are 
            reproduced with similar accuracy}
   \label{fig_predicted_arcs_II}  
\end{figure*} 

Using our solution (masses and source positions) for the case of SL+WL and 8000 iterations in section 
\ref{sect_results}, we can predict the position of the multiple images for each source. In most cases, the 
agreement between these positions and the observed arcs is very good with typical errors of less than $5\arcsec$ 
in the image plane but there are some deviations between the model and the data set that 
are interesting to explore in more detail. Disagreements between the predicted arcs and input data might reveal a 
systematic bias in the solution in that particular region of the image plane or even some tension between the 
identification of the multiple images in the data set. Also, new images 
(from a given system) might be identified with the new model. Figure \ref{fig_predicted_arcs} shows two of the most 
extreme cases where the disagreement between the prediction and the input data is more obvious. In dark grey we show 
the distance to the source position when that particular point in the image plane is projected back into the source 
plane at the redshift of the source. In white we show the observed position of the arcs in the original data set. 
We show the cases for systems 12 and 22.
In system 12 we observe that the model predicts a new image on the 
right side of the field of view. Exploring this position in the original ACS image, 
the alleged new image is easily identified about $10\arcsec$ north of the predicted position. 
Given the fact that this is near a critical curve, $10\arcsec$ are actually a relatively short distance 
along a critical curve. That new image was already correctly identified by \cite{Limousin2007}. 
On the other hand, system 22 shows a clear prediction that is missing from our data set (and others in the literature). 
ACS data shows nothing that resembles this bright and distinctive source indicating that this is either a region that 
is near the regime where multiple sources merge and disappear (as suggested by the fact that by moving the source 
position a few arcseconds the predicted image disappears) or, maybe more likely, that the mass model is not very well 
constrained in this part of the lens plane. 

In figure \ref{fig_predicted_arcs_II} we show additional examples of systems where the model accurately reproduces 
the positions (and also the extension) of the arcs in the data. Among these systems, we included also system 10, 
which has been redefined in this paper based on the new IR data. The new configuration of system 10 seems to be 
consistent with our model, with the exception of the arclet on the right side that is possibly affected by a nearby 
massive galaxy that our model fails to reproduce with enough accuracy.

%\begin{figure*}  
%   \includegraphics[width=8cm]{figs/PredictedArcs_vs_True_System19_New2.ps}
%   \includegraphics[width=8cm]{figs/PredictedArcs_vs_True_System19_Arc1.ps}
%   \caption{Left. Zoomed region around one of the positions precited by our lens model for system 19 (this region is shown as a white square in figure 
%            \ref{fig_predicted_arcs}. Right shows one of the arcs in system 19 (the one marked with a white circle)}
%   \label{fig_predicted_arcs_system19}  
%\end{figure*} 

From our reference model we have also identified a set of 11 new system candidates never published before. These 
new system candidates are listed in table 2 (in the appendix) with IDs ranging from 51 to 61. For all of these 
systems we have assumed a redshift of $z=2$ so the subset of new system candidates is naturally biased to have 
redshifts around this value. The stamps for the arclets in the new system candidates are 
shown in figures \ref{fig_StampsI} and \ref{fig_StampsII} (also in the appendix). The full collection of stamps 
for the arclets in table 2 can be found online at this website\footnote{http://max.ifca.unican.es/diego/FigsA1689/}.

%%%%%%%%%%%%%%%%%%%%%%%%%%%%%%%%%%%%%%%%%%%%  
\section{Conclusions}  
%%%%%%%%%%%%%%%%%%%%%%%%%%%%%%%%%%%%%%%%%%%%
We have presented a robust estimation of the dark matter distribution in the cluster A1689. We explore the range 
of variability of the solutions and identify a region of minimum variance where the solution is stable against 
changes in the configuration of the data set, the number of iterations, the grid resolution and, assumptions made 
of the fiducial deflection field. We also identify regions 
where the results should be taken with more caution.  Our solution can be used to identify additional strongly 
lensed systems. We identify 11 new systems (candidates) and confirm some of the previous identifications 
like including the contentious system number 12 where our solution correctly predicts the fourth arclet identified in 
\cite{Limousin2007}. 

Even though the WL measurements have a typical sampling scale of $1\arcmin$, through the combination of the SL and 
WL data sets in the same minimization algorithm we manage to improve upon this resolution beyond the Einstein 
radius and be sensitive to smaller scales. 
This allows us to resolve details unseen before in the dark matter distribution around and beyond the Einstein radius, 
some of which have no obvious correlation with the luminous matter.  At larger radii ($r>2$ arcmin) 
the sensitivity to smaller details weakens as the SL data set loses its capability to constrain the 
matter distribution at these distances.

\begin{figure}  
\centerline{ \includegraphics[width=8cm]{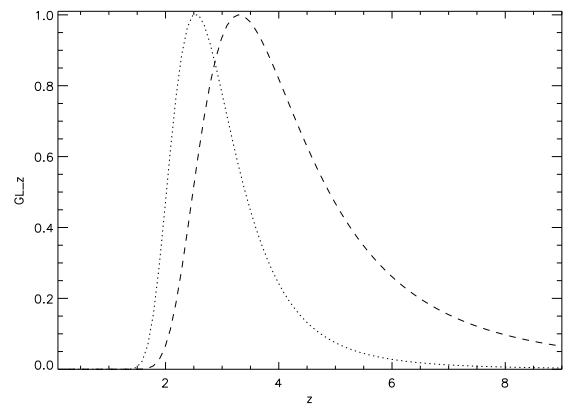}}          
   \caption{Gravitational lensing redshift prediction for systems 8 (dotted) and 9 (dashed) from our reference model.
           }   
   \label{fig_GravLens89}  
\end{figure}

A consequence of knowing the mass distribution of a lens is that it makes it possible to make predictions that 
can be used in other observations. For instance, we can derive redshifts for sources with unknown redshift by 
projecting the system back at different redshifts and finding the redshift at which the system {\it come into focus}. 
Figure  \ref{fig_GravLens89} is an example for two of the photometric redshift systems in our sample. 
For these sources, the lens predicts lower redshifts than the photo-z although the predicted redshift is still 
consistent with the photo-z. 

The mass model can be also used to impose constraints on the cosmological model, for instance through the 
$f_k$ function as shown in section \ref{sect_cosmo} although we also show that in order to get competitive 
constraints one should probably rely on stacking results from multiple clusters.

Finally, our free-form model allows us to determine the mass-to-light ratio of the main galaxies in the cluster. 
We find ratios that are generally consistent with earlier results found in the literature. However, with the 
exception of the central galaxy, the galaxies in our model assume the same $M/L_B$ ratio limiting somehow  
the power of our study. A more detailed study where member galaxies are allowed to take on individual $M/L_B$  
ratios will be considered in a future work.

%%%%%%%%%%%%%%%%%%%%%%%%%%%  
\section{Acknowledgments}  
%%%%%%%%%%%%%%%%%%%%%%%%%%%  
J.M.D acknowledges support of the consolider project CAD2010-00064 and 
AYA2012-39475-C02-01 funded by the Ministerio de Economia y Competitividad. 
JMD also acknowledges the hospitality of the Department of Physics and Astronomy at UPenn during 
part of this research. 
K.U. acknowledges partial support from the National Science Council of Taiwan (grant NSC100-2112-M-001-008-MY3).
MS acknowledges financial support from the agreement ASI/INAF/023/12/0.

 %%%%%%%%%%%%%%%%%%%%%%%%%%%%%%%%%%%%%%%%%%%%%%%%%%%%%%%%%%%%%%%%%%%%%%%  

\bsp  
\label{lastpage}

\bibliographystyle{mn2e}% style aa.bst 
\bibliography{MyBiblio} % your references Yourfile.bib 

\appendix

\section{Compilation of arc positions and Stamps}

    \begin{table*}
    \begin{minipage}{115mm}                                               
    \caption{Full strong lensing data set. The second column shows the new system ID following the original notation 
             of \citep{Broadhurst05a}. The third column shows the original notation \citep{Broadhurst05a}. Systems
             not present in the original paper are left blank in this column. 
            The fourth column indicates previous papers in the literature where that system was identified, 
             B05 is for \citep{Broadhurst05a}, L07 for \citep{Limousin2007}, C10 for \citep{Coe2010} 
             and D14 for the present paper. Fifth and sixths columns show the coordinates of each arclet. 
             Discrepancies with some of the positions published in \citep{Coe2010} have been resolved by 
             (D. Coe, private communication) and are corrected in the present version of the table. 
             The seventh column includes the redshifts used in our study. A negative sign indicates that they 
             are photometric redshifts. The last column shows the $\Delta \beta$ derived from 
             our reference model, see equation \ref{eq_chi2} }
 \label{tab1}
 \begin{tabular}{cccccccc}   
  i &   ID  & B05 &  REF & RAJ2000(h:m:s) & DECJ2000(d:m:s) &  z   &  $\Delta \beta$    \\
  1 &   1.1   & 1.1 & B05 &  13:11:26.257  &  -1:19:58.753   & 3.04 &  1.03  \\  
  2 &   1.2   & 1.2 & B05 &  13:11:26.088  &  -1:20:02.261   & 3.04 &  0.73  \\     
  3 &   1.3   & 1.3 & B05 &  13:11:29.584  &  -1:21:09.475   & 3.04 &  2.50  \\     
  4 &   1.4   & 1.4 & B05 &  13:11:32.870  &  -1:20:29.403   & 3.04 &  1.27  \\     
  5 &   1.5   & 1.5 & B05 &  13:11:31.742  &  -1:20:07.998   & 3.04 &  3.98  \\     
  6 &   1.6   & 1.6 & B05 &  13:11:29.661  &  -1:20:40.413   & 3.04 &  2.44  \\     
  7 &   2.1   & 2.1 & B05 &  13:11:26.331  &  -1:19:57.450   & 2.53 &  0.92  \\  
  8 &   2.2   & 2.2 & B05 &  13:11:32.771  &  -1:20:27.494   & 2.53 &  1.78  \\     
  9 &   2.3   & 2.3 & B05 &  13:11:31.780  &  -1:20:09.147   & 2.53 &  3.42  \\     
 10 &   2.4   & 2.4 & B05 &  13:11:29.619  &  -1:21:08.008   & 2.53 &  2.32  \\     
 11 &   2.5   & 2.5 & B05 &  13:11:29.686  &  -1:20:41.365   & 2.53 &  2.02  \\     
 12 &   3.1   & 3.1 & B05 &  13:11:31.850  &  -1:20:29.520   &-5.47 &  2.30  \\  
 13 &   3.2   & 3.2 & B05 &  13:11:31.979  &  -1:20:35.287   &-5.47 &  1.29  \\     
 14 &   3.3   & 3.3 & B05 &  13:11:31.492  &  -1:20:58.040   &-5.47 &  3.33  \\     
 15 &   4.1   & 4.1 & B05 &  13:11:31.978  &  -1:20:59.355   & 1.10 &  0.18  \\  
 16 &   4.2   & 4.2 & B05 &  13:11:30.326  &  -1:21:14.026   & 1.10 &  2.24  \\     
 17 &   4.3   & 4.3 & B05 &  13:11:30.565  &  -1:20:10.322   & 1.10 &  3.02  \\     
 18 &   4.4   & 4.4 & B05 &  13:11:26.094  &  -1:20:37.422   & 1.10 &  0.94  \\     
 19 &   4.5   & 4.5 & B05 &  13:11:29.653  &  -1:20:31.357   & 1.10 &  2.23  \\     
 20 &   5.1   & 5.1 & B05 &  13:11:28.873  &  -1:20:50.776   & 2.60 &  2.10  \\  
 21 &   5.2   & 5.2 & B05 &  13:11:29.032  &  -1:20:46.153   & 2.60 &  3.10  \\     
 22 &   5.3   & 5.3 & B05 &  13:11:33.927  &  -1:20:22.919   & 2.60 &  5.20  \\     
 23 &   6.1   & 6.1 & B05 &  13:11:30.555  &  -1:19:39.995   & 1.10 &  1.57  \\  
 24 &   6.2   & 6.2 & B05 &  13:11:33.154  &  -1:20:14.174   & 1.10 &  2.70  \\     
 25 &   6.3   & 6.3 & B05 &  13:11:32.558  &  -1:19:56.506   & 1.10 &  1.45  \\     
 26 &   6.4   & 6.4 & B05 &  13:11:32.289  &  -1:20:00.857   & 1.10 &  2.72  \\     
 27 &   7.1   & 7.1 & B05 &  13:11:25.256  &  -1:20:53.843   & 4.87 &  7.68  \\  
 28 &   7.2   & 7.2 & B05 &  13:11:30.478  &  -1:20:15.902   & 4.87 &  3.18  \\     
 29 &   7.3   & 7.3 & B05 &  13:11:29.627  &  -1:20:26.870   & 4.87 &  9.33  \\     
 30 &   8.1   & 8.1 & B05 &  13:11:32.105  &  -1:20:52.909   &-2.67 &  1.32  \\  
 31 &   8.2   & 8.2 & B05 &  13:11:31.210  &  -1:21:07.541   &-2.67 &  3.41  \\     
 32 &   8.3   & 8.3 & B05 &  13:11:31.313  &  -1:20:16.078   &-2.67 &  2.57  \\     
 33 &   8.4   & 8.4 & B05 &  13:11:25.337  &  -1:20:22.162   &-2.67 &  2.87  \\     
 34 &   8.5   & 8.5 & B05 &  13:11:30.136  &  -1:20:32.494   &-2.67 &  6.61  \\     
 35 &   9.1   & 9.1 & B05 &  13:11:30.115  &  -1:19:50.652   &-5.16 &  7.14  \\  
 36 &   9.2   & 9.2 & B05 &  13:11:33.328  &  -1:20:52.335   &-5.16 &  1.31  \\     
 37 &   9.3   & 9.3 & B05 &  13:11:28.554  &  -1:21:17.805   &-5.16 &  2.08  \\     
 38 &   9.4   & 9.4 & B05 &  13:11:26.079  &  -1:20:28.927   &-5.16 &  4.37  \\     
 39 &  10.1   &10.1 & B05 &  13:11:33.786  &  -1:20:52.855   & 1.83 &  1.78  \\  
 40 &  10.2   &10.2 & B05 &  13:11:27.857  &  -1:20:14.477   & 1.83 &  3.89  \\     
 41 &  10.3   &10.3 & B05 &  13:11:29.125  &  -1:20:29.744   & 1.83 &  2.67  \\     
 42\footnote{System has been re-organized} &  10.4   &12.2 & B05 &  
13:11:27.166  &  -1:20:56.946   & 1.83 &  3.72 \\
 43\footnote{System has been re-organized} &  10.5   &12.3 & B05 &  
13:11:27.033  &  -1:20:53.910   & 1.83 &  4.57 \\
 44 &  11.1   &11.1 & B05 &  13:11:33.149  &  -1:21:08.754   & 2.50 &  2.35  \\  
 45 &  11.2   &11.2 & B05 &  13:11:28.866  &  -1:20:03.292   & 2.50 &  1.54  \\     
 46 &  11.3   &11.3 & B05 &  13:11:29.300  &  -1:20:28.381   & 2.50 &  2.05  \\     
 47 &  12.1   &12.1 & B05 &  13:11:30.171  &  -1:19:53.471   & 1.82 &  6.34  \\  
 48 &  12.2   &12.4 & B05 &  13:11:28.771  &  -1:21:12.265   & 1.82 &  2.66  \\     
 49\footnote{System has been re-organized. New arclet candidate} &  12.3   &31.2 & C10 &  
13:11:33.081  &  -1:20:46.390   & 1.82 &  1.31 \\
 \end{tabular}
 \end{minipage}
\end{table*}

\setcounter{table}{0}
    \begin{table*}
    \begin{minipage}{115mm}                                               
    \caption{cont.}
% \label{tab1}
 \begin{tabular}{cccccccc} 
  i &   ID  & B05 &  REF & RAJ2000(h:m:s) & DECJ2000(d:m:s) &  z   &  $\Delta \beta$    \\
 50\footnote{New arclet candidate not used in our primary analysis} &  12.4   &31.4 & C10 &  
13:11:26.303  &  -1:20:24.080   & 1.82 &  2.64 \\
 51 &  13.1   &13.1 & B05 &  13:11:32.631  &  -1:19:26.371   &-1.02 &  1.05  \\  
 52 &  13.2   &13.2 & B05 &  13:11:32.795  &  -1:19:27.831   &-1.02 &  0.97  \\     
 53 &  13.3   &13.3 & B05 &  13:11:33.200  &  -1:19:33.134   &-1.02 &  2.03  \\     
 54 &  14.1   &14.1 & B05 &  13:11:28.835  &  -1:21:43.802   & 3.40 &  0.65  \\  
 55 &  14.2   &14.2 & B05 &  13:11:29.266  &  -1:21:44.623   & 3.40 &  0.65  \\     
 56 &  15.1   &15.1 & B05 &  13:11:27.882  &  -1:20:17.196   & 1.80 &  0.50  \\  
 57 &  15.2   &15.2 & B05 &  13:11:33.883  &  -1:20:53.311   & 1.80 &  1.03  \\     
 58 &  15.3   &15.3 & B05 &  13:11:29.046  &  -1:20:29.573   & 1.80 &  0.96  \\     
 59 &  16.1   &16.1 & B05 &  13:11:27.790  &  -1:20:27.319   &-2.01 &  1.63  \\  
 60 &  16.2   &16.2 & B05 &  13:11:28.721  &  -1:20:30.546   &-2.01 &  3.14  \\     
 61 &  16.3   &16.3 & B05 &  13:11:34.205  &  -1:20:48.402   &-2.01 &  1.55  \\     
 62 &  17.1   &17.1 & B05 &  13:11:30.463  &  -1:20:26.890   & 2.60 &  3.04  \\  
 63 &  17.2   &17.2 & B05 &  13:11:30.196  &  -1:20:29.765   & 2.60 &  6.11  \\     
 64 &  17.3   &17.3 & B05 &  13:11:24.787  &  -1:20:43.865   & 2.60 &  9.12  \\     
 65 &  18.1   &18.1 & B05 &  13:11:28.052  &  -1:20:11.540   & 1.80 &  1.23  \\  
 66 &  18.2   &18.2 & B05 &  13:11:33.627  &  -1:20:56.539   & 1.80 &  0.86  \\     
 67 &  18.3   &18.3 & B05 &  13:11:29.169  &  -1:20:29.392   & 1.80 &  0.43  \\     
 68 &  19.1   &19.1 & B05 &  13:11:31.440  &  -1:20:24.597   & 2.60 &  3.53  \\  
 69 &  19.2   &19.2 & B05 &  13:11:25.047  &  -1:20:22.003   & 2.60 &  5.47  \\     
 70 &  19.3   &19.3 & B05 &  13:11:31.762  &  -1:21:01.315   & 2.60 &  3.61  \\     
 71 &  19.4   &19.4 & B05 &  13:11:31.859  &  -1:20:59.131   & 2.60 &  2.02  \\     
 72 &  19.5   &19.5 & B05 &  13:11:30.017  &  -1:20:35.961   & 2.60 &  3.05  \\     
 73 &  21.1   &21.1 & B05 &  13:11:30.833  &  -1:20:47.776   &-1.78 &  1.45  \\  
 74 &  21.2   &21.2 & B05 &  13:11:30.608  &  -1:20:46.743   &-1.78 &  2.42  \\     
 75 &  21.3   &21.3 & B05 &  13:11:25.061  &  -1:20:13.207   &-1.78 &  3.51  \\     
 76 &  22.1   &22.1 & B05 &  13:11:29.493  &  -1:20:10.794   & 1.70 &  3.73  \\  
 77 &  22.2   &22.2 & B05 &  13:11:29.423  &  -1:20:25.762   & 1.70 &  1.71  \\     
 78 &  22.3   &22.3 & B05 &  13:11:32.222  &  -1:21:17.917   & 1.70 &  4.69  \\     
 79 &  23.1   &23.1 & B05 &  13:11:29.337  &  -1:20:12.016   &-2.00 &  2.94  \\  
 80 &  23.2   &23.2 & B05 &  13:11:29.361  &  -1:20:24.891   &-2.00 &  1.06  \\     
 81 &  23.3   &23.3 & B05 &  13:11:32.465  &  -1:21:17.199   &-2.00 &  3.45  \\     
 82 &  24.1   &24.1 & B05 &  13:11:28.998  &  -1:20:58.177   & 2.60 &  0.06  \\  
 83 &  24.2   &24.2 & B05 &  13:11:31.871  &  -1:19:52.560   & 2.60 &  0.95  \\     
 84 &  24.3   &24.3 & B05 &  13:11:30.101  &  -1:19:36.140   & 2.60 &  3.12  \\     
 85 &  24.4   &24.4 & B05 &  13:11:33.525  &  -1:20:21.863   & 2.60 &  3.32  \\     
 86 &  24.5   &24.5 & B05 &  13:11:29.436  &  -1:20:38.999   & 2.60 &  7.13  \\      
 87\footnote{Original arclet not used in our analysis} &  25.1   &25.1 & B05 &  
13:11:28.302  &  -1:20:36.990   & 2.50 &  9.30  \\
 88\footnote{Original arclet not used in our analysis. Multiple possibilities for this arclet.} &  
25.2   &25.2 & B05 &  13:11:34.455  &  -1:20:35.581   & 2.50 &  6.71  \\
 89\footnote{New arclet not used in our primary analysis. Multiple possibilities for this arclet.} &  
25.2   &56.1 &  D14 &  13:11:33.970  &  -1:20:41.300   & 2.50 & 22.47  \\
 90\footnote{New arclet not used in our primary analysis. Multiple possibilities for this arclet.} &  
25.2   &45.2 &  C10 &  13:11:35.489  &  -1:20:32.950   & 2.50 & 37.17  \\
 91 &  28.1   &28.1 & B05 &  13:11:28.105  &  -1:20:12.907   &-5.45 &  0.83  \\  
 92 &  28.2   &28.2 & B05 &  13:11:34.067  &  -1:21:02.009   &-5.45 &  2.62  \\     
 93\footnote{New arclet not used in our primary analysis} &  28.3   &28.3 & C10 &  
13:11:29.100  &  -1:20:28.610   &-5.45 &  2.11  \\
 94 &  29.1   &29.1 & B05 &  13:11:29.033  &  -1:20:59.909   & 2.50 &  1.04  \\  
 95 &  29.2   &29.2 & B05 &  13:11:29.845  &  -1:19:36.215   & 2.50 &  3.95  \\     
 96 &  29.3   &29.3 & B05 &  13:11:31.952  &  -1:19:54.565   & 2.50 &  1.44  \\     
 97 &  29.4   &29.4 & B05 &  13:11:33.433  &  -1:20:22.815   & 2.50 &  3.72  \\     
 98 &  29.5   &29.5 & B05 &  13:11:29.537  &  -1:20:38.603   & 2.50 &  9.75  \\     
 99 &  30.1   &30.1 & B05 &  13:11:32.228  &  -1:19:21.826   & 3.00 &  1.95  \\  
100 &  30.2   &30.2 & B05 &  13:11:32.990  &  -1:19:28.069   & 3.00 &  0.49  \\     
101 &  30.3   &30.3 & B05 &  13:11:33.461  &  -1:19:34.691   & 3.00 &  2.44  \\     
102 &  32.1   &   & C10 &  13:11:31.998  &  -1:20:05.530   & 3.00 &  3.69  \\  
103 &  32.2   &   & C10 &  13:11:33.023  &  -1:20:22.900   & 3.00 &  1.10  \\     
 \end{tabular}
 \end{minipage}
\end{table*}

\setcounter{table}{0}
    \begin{table*}
    \begin{minipage}{115mm}                                               
    \caption{cont.}
% \label{tab1}
 \begin{tabular}{cccccccc}    
  i &   ID    & B05 &  REF & RAJ2000(h:m:s) & DECJ2000(d:m:s) &  z   &  $\Delta \beta$    \\
104 &  32.3   &   & C10 &  13:11:29.396  &  -1:21:04.830   & 3.00 &  0.12  \\     
105 &  32.4   &   & C10 &  13:11:29.611  &  -1:20:45.350   & 3.00 &  3.53  \\     
106\footnote{Multiple possibilities for this arclet} &  32.5   &   & L07 &  
13:11:26.407  &  -1:19:59.600   & 3.00 &  8.00 \\
107\footnote{Multiple possibilities for this arclet} &  32.5   &   & D14 &  
13:11:27.470  &  -1:19:41.520   & 3.00 &  6.86 \\
108 &  33.1   &  &  C10 &  13:11:28.256  &  -1:21:02.660   & 4.58 &  7.29  \\  
109 &  33.2   &  &  C10 &  13:11:34.460  &  -1:20:35.600   & 4.58 &  7.29  \\     
110 &  35.1   &  &  C10 &  13:11:28.367  &  -1:21:01.350   & 1.90 &  2.45  \\  
111 &  35.2   &  &  C10 &  13:11:33.765  &  -1:20:34.520   & 1.90 &  4.12  \\     
112 &  35.3   &  &  C10 &  13:11:29.238  &  -1:20:36.660   & 1.90 &  5.60  \\     
113 &  36.1   &  &  C10 &  13:11:31.373  &  -1:19:47.940   & 3.00 &  0.26  \\  
114 &  36.2   &  &  C10 &  13:11:31.493  &  -1:19:49.390   & 3.00 &  0.26  \\     
115 &  40.1   &  &  C10 &  13:11:30.067  &  -1:20:14.040   & 2.52 &  4.98  \\  
116 &  40.2   &  &  C10 &  13:11:25.983  &  -1:21:05.290   & 2.52 &  4.98  \\     
117 &  41.1   &  &  C10 &  13:11:27.679  &  -1:20:50.710   &-2.50 & 12.23  \\  
118 &  41.2   &  &  C10 &  13:11:35.329  &  -1:20:31.380   &-2.50 & 29.37  \\     
119 &  41.3   &  &  C10 &  13:11:28.769  &  -1:20:36.120   &-2.50 & 17.19  \\     
120 &  42.1   &  &  C10 &  13:11:28.479  &  -1:19:44.980   &-2.00 &  4.31  \\  
121 &  42.2   &  &  C10 &  13:11:31.077  &  -1:19:55.520   &-2.00 &  5.44  \\     
122 &  42.3   &  &  C10 &  13:11:33.317  &  -1:20:37.890   &-2.00 &  1.60  \\     
123 &  42.4   &  &  C10 &  13:11:28.842  &  -1:21:09.490   &-2.00 &  1.22  \\     
124 &  44.1   &  &  C10 &  13:11:28.324  &  -1:20:23.380   &-2.00 & 11.81  \\  
125 &  44.2   &  &  C10 &  13:11:34.338  &  -1:21:04.010   &-2.00 & 11.81  \\     
126 &  46.1   &  &  C10 &  13:11:31.476  &  -1:20:49.190   &-2.50 &  4.65  \\  
127 &  46.2   &  &  C10 &  13:11:24.766  &  -1:20:15.980   &-2.50 &  4.65  \\     
128 &  48.1   &  &  C10 &  13:11:31.365  &  -1:20:38.470   &-2.00 &  2.29  \\  
129 &  48.2   &  &  C10 &  13:11:24.911  &  -1:20:19.820   &-2.00 &  2.29  \\     
130 &  49.1   &  &  C10 &  13:11:28.660  &  -1:20:15.470   &-2.00 &  2.76  \\  
131 &  49.2   &  &  C10 &  13:11:33.376  &  -1:21:08.730   &-2.00 &  2.76  \\     
132 &  50.1   &  &  C10 &  13:11:32.387  &  -1:20:45.600   &-2.50 &  2.16  \\  
133 &  50.2   &  &  C10 &  13:11:30.828  &  -1:21:11.080   &-2.50 &  2.13  \\     
134 &  50.3   &  &  C10 &  13:11:31.467  &  -1:20:15.660   &-2.50 &  0.05  \\     
135 &  51.1   &  &  D14 &  13:11:33.650  &  -1:20:17.710   &-2.00 &  3.05  \\  
136 &  51.2   &  &  D14 &  13:11:30.470  &  -1:19:34.740   &-2.00 &  2.09  \\     
137 &  51.3   &  &  D14 &  13:11:32.050  &  -1:19:45.140   &-2.00 &  1.51  \\     
138 &  52.1   &  &  D14 &  13:11:29.480  &  -1:19:35.920   &-1.80 &  1.10  \\  
139 &  52.2   &  &  D14 &  13:11:33.090  &  -1:20:17.170   &-1.80 &  1.10  \\     
140 &  53.1   &  &  D14 &  13:11:31.500  &  -1:20:06.590   &-3.00 &  1.72  \\  
141 &  53.2   &  &  D14 &  13:11:32.920  &  -1:20:36.490   &-3.00 &  5.26  \\     
142 &  53.3   &  &  D14 &  13:11:30.120  &  -1:21:21.210   &-3.00 & 12.79  \\     
143 &  53.4   &  &  D14 &  13:11:25.910  &  -1:20:08.430   &-3.00 &  6.02  \\  
144 &  54.1   &  &  D14 &  13:11:31.480  &  -1:20:09.980   &-2.00 &  1.75  \\  
145 &  54.2   &  &  D14 &  13:11:32.690  &  -1:20:37.270   &-2.00 &  1.34  \\     
146 &  54.3   &  &  D14 &  13:11:29.900  &  -1:21:12.120   &-2.00 &  2.66  \\     
147 &  54.4   &  &  D14 &  13:11:25.770  &  -1:20:11.550   &-2.00 &  4.04  \\     
148 &  55.1   &  &  D14 &  13:11:30.920  &  -1:20:18.690   &-1.50 &  2.04  \\  
149 &  55.2   &  &  D14 &  13:11:25.480  &  -1:20:31.180   &-1.50 &  2.04  \\     
150 &  56.1   &  &  D14 &  13:11:28.558  &  -1:19:43.790   &-2.00 &  5.36  \\  
151 &  56.2   &  &  D14 &  13:11:31.366  &  -1:19:59.450   &-2.00 &  5.03  \\     
152 &  56.3   &  &  D14 &  13:11:33.165  &  -1:20:33.110   &-2.00 &  1.69  \\     
153 &  56.4   &  &  D14 &  13:11:29.049  &  -1:21:06.660   &-2.00 &  2.17  \\     
154 &  57.1   &  &  D14 &  13:11:31.206  &  -1:19:54.500   &-2.00 &  3.72  \\  
155 &  57.2   &  &  D14 &  13:11:33.326  &  -1:20:33.100   &-2.00 &  2.55  \\     
156 &  57.3   &  &  D14 &  13:11:28.823  &  -1:21:05.100   &-2.00 &  2.52  \\     
157 &  57.4   &  &  D14 &  13:11:29.953  &  -1:19:43.780   &-2.00 &  1.72  \\     
158 &  58.1   &  &  D14 &  13:11:33.165  &  -1:20:03.730   &-2.00 &  6.87  \\  
159 &  58.2   &  &  D14 &  13:11:29.351  &  -1:20:51.190   &-2.00 &  6.87  \\     
160 &  59.1   &  &  D14 &  13:11:26.920  &  -1:20:39.240   &-2.00 &  1.19  \\  
161 &  59.2   &  &  D14 &  13:11:27.136  &  -1:20:48.140   &-2.00 &  1.19  \\     
162 &  60.1   &  &  D14 &  13:11:29.988  &  -1:20:19.280   &-2.00 &  0.51  \\  
163 &  60.2   &  &  D14 &  13:11:29.893  &  -1:20:23.050   &-2.00 &  0.51  \\     
164 &  61.1   &  &  D14 &  13:11:32.277  &  -1:21:22.900   &-2.00 &  7.83  \\  
165 &  61.2   &  &  D14 &  13:11:29.524  &  -1:20:10.240   &-2.00 &  7.83  \\     
 \end{tabular}
 \end{minipage}
\end{table*}

\begin{figure*}  
\centerline{ \includegraphics[width= 18cm]{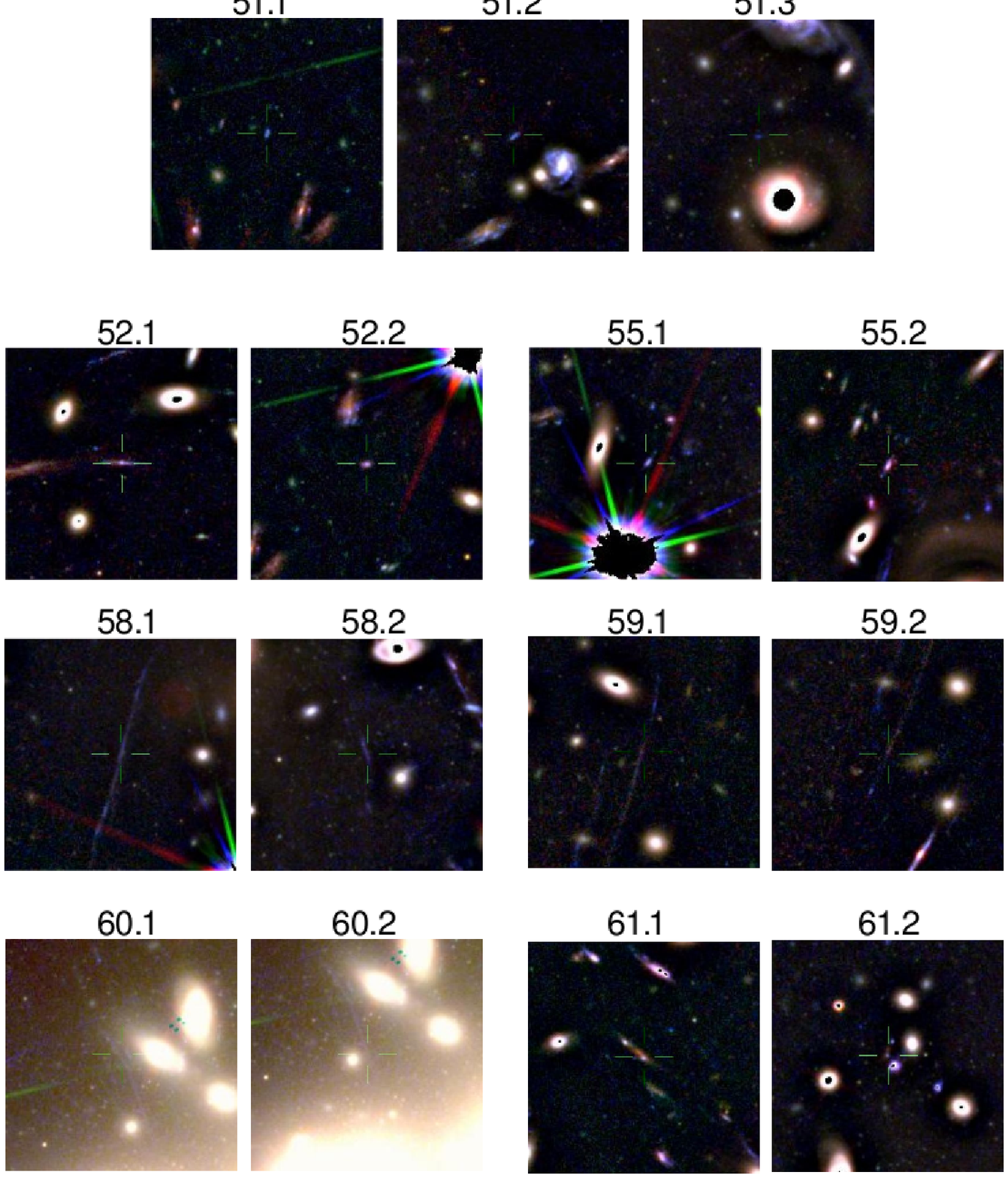}}          
%   \caption{RGB composite image of 3 Hubble bands (814w 475w and nir).    }   
   \caption{Images have been filtered to reduce light glare from member 
            galaxies except for system 60 which is shown in its original 
            unfiltered version.}
   \label{fig_StampsI}  
\end{figure*}  

\begin{figure*}  
\centerline{ \includegraphics[width=18cm]{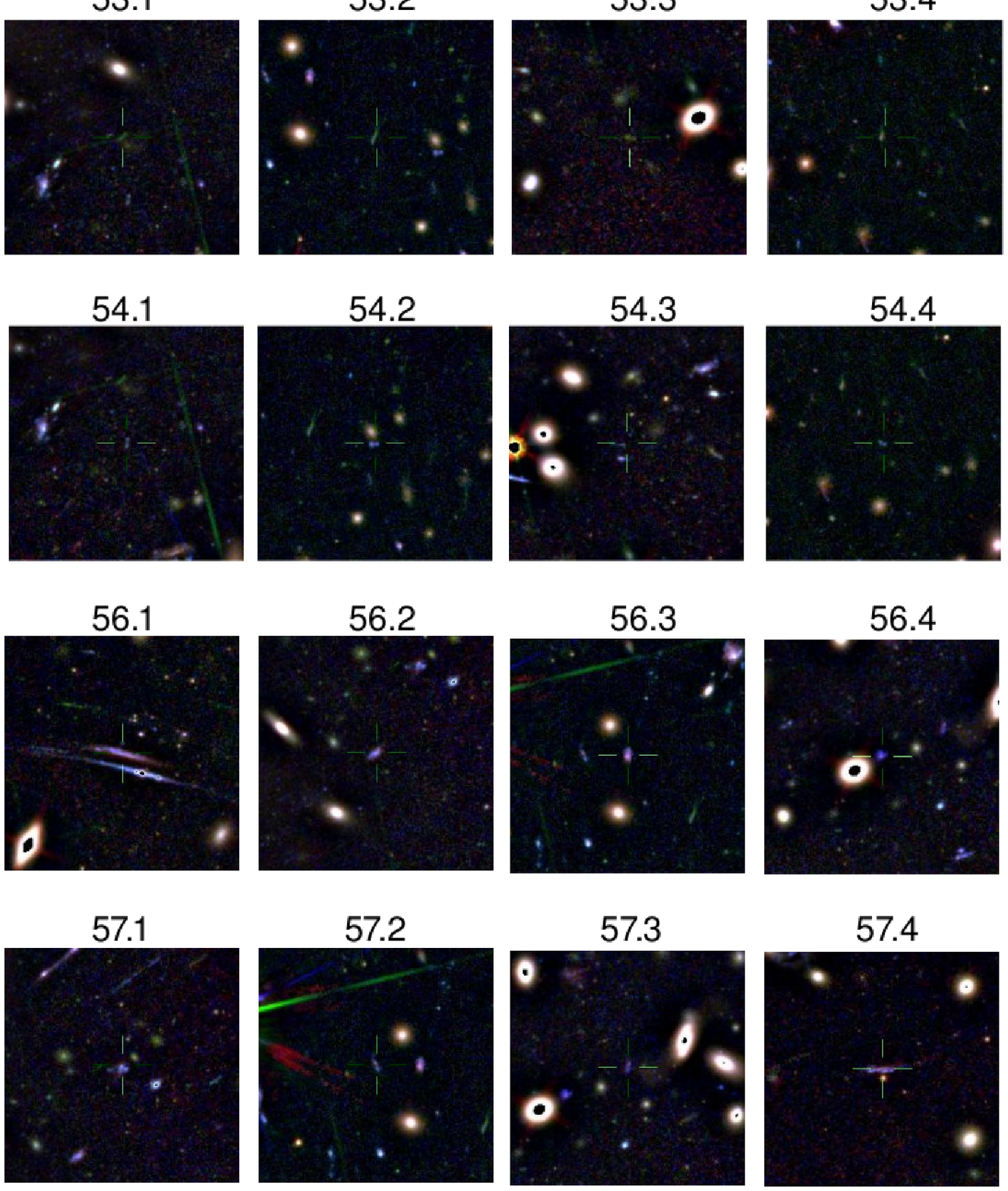}}          
   \caption{Light from member galaxies has been reduced through a high-pass 
            filter.}
   \label{fig_StampsII}  
\end{figure*}

\end{document}